\RequirePackage{fix-cm}
\documentclass[smallextended]{svjour3}       
\smartqed  
\usepackage{graphicx}
\usepackage{bm}
\usepackage{epsfig}
\newcommand{\beq}{\begin{equation}}
\newcommand{\eeq}{\end{equation}}
\newcommand{\beqa}{\begin{eqnarray}}
\newcommand{\eeqa}{\end{eqnarray}}

 \journalname{General Relativity and Gravitation}
\begin{document}

\title{Axial and polar modes for the ring down of
a Schwarzschild black hole with an $r$ dependent 
mass-function
\thanks{DGAPA-PAPIIT IN100418}
}

\titlerunning{Axial and polar ring down modes}        

\author{Peter O. Hess         \and
        Enrique L\'opez-Moreno 
}


\institute{P. O. Hess \at
              Instituto de Ciencias Nucleares, Universidad Nacional 
Aut\'onoma de M\'exico, Circuito Exterior, C.U., 
A.P. 70-543, 04510 M\'exico D.F., Mexico \\
and \\
Frankfurt Institute for Advanced Studies, Johann Wolfgang 
Goethe Universit\"at, Ruth-Moufang-Str. 1, 60438 Frankfurt am Main, Germany
              \email{hess@nucleares.unam.mx}           
           \and
           E. L\'opez-Moreno \at
              Facultad de Ciencias, Universidad Nacional Aut\'onoma 
de M\'exico, Mexico-City, Mexico
}

\date{Received: date / Accepted: date}

\maketitle

\begin{abstract}
The axial and polar modes for the ring down of a 
Schwarzschild black hole
are calculated, by first deriving the Regge-Wheeler and Zerilli equations, respectively, and finally 
applying the {Asymptotic Iteration Method}
(AIM). We were able to reach up to 500 iterations, obtaining
for the first time
convergence for a wide range of large damping modes.
The {\it General Relativity} (GR) and a particular 
version of an extended model 
with an $r$-dependent mass-function are compared. 
This mass-function allows an analytical solution for
the Tortoise coordinate. 
The example of the mass-function corresponds to 
the leading correction for extended theories and 
serves as a starting point
to treat other $r$-dependent parameter 
mass-functions.
\end{abstract}

\section{Introduction}
\label{intro}

{\it General Relativity} (GR) is one of the best tested theories,
which accounts for the observations in the solar system
\cite{will} and also its prediction of gravitational
waves \cite{maggiore} was confirmed recently 
\cite{abbot1,abbot2}. The first indirect proof of these 
waves stems
from the 1970's \cite{hulse}, through the observation of changes
in the orbital frequency of a neutron star binary. 
The black hole merger consists
of two phases: the inspiral and the ring-down phase. 
For the
description of the inspiral phase the result depends 
very much on the approximation used \cite{maggiore,hess-2016} or 
on a non-linear  
hydro-dynamic approach \cite{rezzolla-book}.
For the ring-down phase the situation is "simpler",
because only the stability of the black hole under metric
perturbations has to be studied. 
The investigation
of the ring-down phase can be traced back to S. 
Chandrasekhar's book on the mathematics of black holes
\cite{chandra} and \cite{chandra1975a,chandra1975b}
(though, not the first). 
The equation for the calculation of
the ring-down frequencies where treated in \cite{zerilli} 
for the polar modes and in \cite{RW} for the axial modes.  
These equations permit to calculate the frequencies, solving the eigenvalue problem of the equation, which is much
simpler than to implement a dynamical theory. There are several 
methods to solve these equations, 
as for example the so-called
{\it Asymptotic Iteration Method} (AIM) 
\cite{ciftci2003,ciftci2005}, with an improved approach 
published in \cite{cho2012}.  

For this reason, we restrict
our analysis to the study of the ring-down modes 
only and in addition
to a non-rotating star, i.e., to the Schwarzschild metric.
We follow closely the method described in the book of S. Chandrasekhar \cite{chandra}, Chapter 4 on the perturbations of a Schwarzschild black hole.

An interesting questions is: What changes,
when a $r$-dependent mass-function $m(r)$ is used,
instead of a constant mass parameter $m_0$? 
There is a $1/r^4$ leading order, due to the following
arguments: a) $1/r^2$ corrections are excluded due to 
observations in the solar system \cite{will}. b)
Also $1/r^3$ corrections are excluded due to adjusting to
the inspiral phase 
\cite{nielsen2018,nielsen2019} in the first observed gravitational event. Thus, the leading corrections are
of the order $1/r^4$. Other corrections, as appearing in 
cosmological models (de Sitter), are not included but the
path explained can be extended to it.
It is probable that the constant mass $m_0$ is
substituted by a function in $r$ as soon as GR is extended
and, therefore, it is interesting to ask what kind
of changes one can expect? Does the $1/r^4$ correction
to the metric still lead to stable modes? 
Is isospectrality between axial and polar modes
maintained? These are 
some of the motivations
of this contribution. To make life easier, we will use
a particular mass-function which still implies an 
event horizon, enabling us to extend directly,
with few modifications, the calculations reported in \cite{chandra}.

Another example is
given in \cite{hess2009,book} where the
{\it pseudo-complex General Relativity} (pc-GR) is proposed,
which adds in the vicinity of a black hole a distribution of
dark energy, which is repulsive and 
from a certain value of the coupling constnat halts the collapse of a star, before forming an event horizon and a singularity. In all applications up to now, 
for practical reasons, the coupling 
constant is chosen such that
there is still an event horizon at $r=\frac{3}{2}m_0$.
First observable predictions are published in
\cite{MNRAS2013,MNRAS2014}.
More recent descriptions of this model can be found in 
\cite{highenergy,universe,PPNP}. One reason for
using this theory becomes obvious in the main
body of the text: The mass-function in pc-GR stars with
a $1/r^4$ correction and 
we show that it permits an analytical
solution for the Tortoise coordinate, thus, it 
provides us with
a controlled handling of the asymptotic 
limit of the solutions.
This property helps to understand the changes in the
spectrum of the {\it Quasinormal Modes} (QNM) in the
ring-down phase of a black hole and its stability.

Because the mass-function is such that there is still
an event-horizon, one can proceed in an analog way as
in GR. The only question is how to treat the accumulation
of dark energy around a black hole. The distribution defines
es new vacuum and it will be shown that it only
depends on the total central mass through
a coupling constant, which is not changed 
when perturbations are included.

The paper is organized as follows: In Section \ref{schwarz} the Schwarzschild limit 
will be discussed. In
Section \ref{axial-RW} the easier to
 treat case of axial modes
are determined and in Section \ref{polarmodes} 
the polar modes.
In Section \ref{zerr} the Regge-Wheeler and Zerilli equations
are derived and the numerical method to solve the differential
equation is shortly explained.
In Section \ref{ZE-RAIM}
the asymptotic limit of the corresponding
solution is calculated, using an 
{\it analytical solution} for
the Tortoise coordinate. 
The spectrum of axial and polar modes are determined,
within GR and its extension.
We will show that the isospectral symmetry observed in GR, namely
that the frequencies of the axial and polar modes
are the same, is not maintained for an 
$r$-dependent mass-function, 
though, some similar structures are present.
In Section \ref{concl} the Conclusions are drawn.

\section{The Schwarzschild solution}
\label{schwarz}

Following the notation of 
\cite{chandra,chandra1975a,chandra1975b},
the length element is given by

\beqa
ds^2 & = & e^{2\nu}(dt)^2 - e^{2\psi} \left( d\phi - \omega dt - q_2 dx_2 -q_3 dx_3\right)^2 
\nonumber \\
&&
- e^{2\mu_2}(dx_2)^2 - e^{2\mu_3}
(dx_3)^2
~~~,
\label{ds2}
\eeqa
where $dx_2 = r$ and $dx_3 =\theta$, the azimuth angle. The functions $\nu$, $\phi$, $\omega$, $q_2$, $q_3$, 
$\mu_2$ and $\mu_3$ depend, in general, 
on $t$, $x_2=r$ and $x_3=\theta$,
though, in this contribution we restrict to
a spherical symmetry. The components of the Ricci
and the Einstein tensor can be retrieved from the 
book by S. Chandrasekhar \cite{chandra}, in
terms of the above functions. Care has to be taken
in comparing notations: With the convention of
\cite{chandra}, the tensor
components of thr Riemann and Einstein tensor
are all at the lower position, but 
by directly calculating these components, one can show
that they are equivalent to $R^\mu_{~\nu}$
and $G^\mu_{~\nu}$, which is also explained in
\cite{chandra}. This is important, as will be seen
further below. 

The {\it metric perturbations} are introduced around the 
generalized Schwarzschild metric

\beqa
e^{2\nu} & = & e^{-2\mu_2} ~=~ \left(1-\frac{2 m(r)}{r}\right) ~=~ \frac{\Delta}{r^2} 
\nonumber \\
e^{\mu_3}& = & r ~,~ e^\psi ~=~ r {\rm sin}\theta
\nonumber \\
{\rm and} && 
\nonumber \\
\omega & = & q_2 ~=~ q_3~= ~0 ~,~ \Delta~=~ r^2 - 2m(r) r
~~~.
\label{schw}
\eeqa
The $m(r)$ is the parameter mass-function, which is $m_0$
in the GR case but will depend on $r$ for a generalized form,
as will be used later. It is preferable to 
define the dimensionless
coordinate $y=\frac{r}{m_0}$ and the dimensionless
mass-function $m(y)=\frac{m(r)}{m_0}$ (for simplicity, we use the same letter $m$ for the mass-function). 

Later, we will propose a particular mass-function,
namely $m(y)=\left(1-\frac{b}{6y^3}\right)$.
for $b=\frac{81}{8}$ the $g_{00}$ component of the
metric is zero at $y=\frac{3}{2}$, i.e., it exhibits 
an event horizon. In conclusion, the object under 
study is a black hole with an event horizon at
$\frac{3}{2}m_0$. Thus, the advantage in
using $b=\frac{81}{8}$ is that there is still an event
horizon and considerations can be limited to the outside
region of the black hole.

Using an arbitrary $b$ the Ricci and 
Einstein tensor components satisfy the equations

\beqa
R^\mu_{~\nu} & = & M^{\mu}_{\nu}
~,~ G^\mu_{~\nu} ~=~ 8\pi T^mu_{~\nu}
~~~,
\label{R-and-G-tensors}
\eeqa
with

\beqa
\left( M^\mu_{~\nu}\right) & = &
\left(
\begin{array}{cccc}
-\frac{b}{y^6} & 0 & 0 & 0 \\
0 & -\frac{b}{y^6} & 0 & 0 \\
0 & 0 & \frac{b}{2y^6} & 0 \\
0 & 0 & 0 & \frac{b}{2y^6} \\
\end{array}
\right)
\nonumber \\
8\pi \left(T^\mu_{~\nu} \right) & = & 
\left(
\begin{array}{cccc}
-\frac{b}{2y^6} & 0 & 0 & 0 \\
0 & -\frac{2b}{y^6} & 0 & 0 \\
0 & 0 & \frac{b}{y^6} & 0 \\
0 & 0 & 0 & \frac{b}{y^6} \\
\end{array}
\right)
~=~ 8 \pi
\left(
\begin{array}{cccc}
-\varrho^\Lambda & 0 & 0 & 0 \\
0 & p^\Lambda_r & 0 & 0 \\
0 & 0 & p^\Lambda_\vartheta & 0 \\
0 & 0 & 0 & p^\Lambda_\vartheta \\
\end{array}
\right)
~~~,
\label{R-and-G.tensors-2}
\eeqa
where $\varrho^\Lambda$ is a dark energy density,
$p^\Lambda_r$ the radial pressure and 
$p^\Lambda_\vartheta$ a tangential pressure.
Note that $b$ is an {\it interaction constant}
which couples the amount of dark energy accumulating
near the black hole to the central mass of the black hole.
Thus, as the gravitational constant, the $b$ is 
{\it fixed}, not subject to variation. The $r$-dependence
of the dark energy determines the vacuum structure,
different to the one in GR, but still invariable.

Perturbations are introduced in first order contributions 
for $\delta \omega$, $\delta q_2$, 
$\delta q_3$, $\delta \mu_2$,
$\delta \mu_3$ and $\delta \psi$. Axial waves (negative parity) are related to the perturbations
of $\delta\omega$, $\delta q_2$ and $\delta q_3$, 
while polar waves (positive parity) are related by the perturbations $\delta \nu$, $\delta\mu_2$, 
$\delta\mu_3$ and $\delta\psi$ \cite{chandra}.
When the variation is applied to the components
of the Ricci  and Einstein tensor, considering that
the interaction constant can not be varied because it
only depends on the total central mass which is not
changed. As a consequence,
we have $\delta R^\mu_{~\nu} = 0$ and 
$\delta G^\mu_{~\nu} = 0$,
as in the standard GR \cite{chandra}.

\section{Axial modes: Regge-Wheeler equation}
\label{axial-RW}

To follow this section, please consult the book
of S. Chandrasekhar \cite{chandra}.
The function $\nu (r)$, $\psi (r)$, $\mu_2 (r)$,
$\mu_3 (r)$, $\omega (r)$, $q_2(r)$ and $q_3 (r)$ are
maintained as general functions in $r$.
In \cite{chandra} the Riemann 
and Ricci tensor components are written in terms of these
functions, as are the Einstein tensor components 
$G^{\mu}_{~\nu}$. Only at the very last the explicit
functions are substituted by their expressions in
the Schwarzschild metric, which 
in this contribution is of the form
listed in (\ref{schw}).

Demanding the invariance of these components under variation,
leads to
$\delta R^{1}_{~2}=0$ and $\delta R^{1}_{~3} = 0$
\cite{chandra} (see also discussion in Section
\ref{schwarz}, which in turn results in the equations
(using the same notation as in \cite{chandra})

\beqa
\left( e^{3\psi + \nu - \mu_2 - \mu_3}Q_{23}\right)_{\mid 3} & = & -e^{3\psi - \nu +\mu_3 - \mu_2}Q_{02\mid 0}
\nonumber \\
\left( e^{3\psi + \nu - \mu_2 - \mu_3}Q_{23}\right)_{\mid 2} & = & +e^{3\psi - \nu +\mu_2 - \mu_3}Q_{03\mid 0}
~~~,
\eeqa
where "$\mid k$" denotes the usual derivative with respect to the variable $x_k$ and the $Q_{ab}$ are defined as

\beqa
Q_{ab} & = & q_{a\mid b}-q_{b\mid a} ~,~ Q_{a0} ~=~ a_{a\mid 0} - \omega_{\mid a}
~~~.
\eeqa

With the ansatz

\beqa
Q(t,r,\theta ) & = & \Delta Q_{23} {\rm sin}^3 \theta ~=~ \Delta (q_{2\mid 3} - q_{3\mid 2}){\rm sin}^3 \theta 
~~~,
\label{ansatz1}
\eeqa
using (\ref{schw}),
we arrive at the equations

\beqa
\frac{1}{r^4 {\rm sin}~ 3\theta} \frac{\partial Q}{\partial \theta} & = & -\left( \omega_{\mid 2} - q_{2\mid 0}\right)_{\mid 0}
\nonumber \\
\frac{\Delta}{r^4 {\rm sin}~ 3\theta} \frac{\partial Q}{\partial r} & = & +\left( \omega_{\mid 3} - q_{3\mid 0}\right)_{\mid 0}
~~~.
\eeqa

Eliminating $\omega$ and assuming a time 
dependence of $e^{i\omega t}$, we arrive at

\beqa
r^4 \frac{\partial}{\partial r} \left( \frac{\Delta}{r^4} \frac{\partial Q}{\partial r}\right) + {\rm sin}^3\theta 
\frac{\partial}{\partial \theta} \left( \frac{1}{{\rm sin}^3 \theta} \frac{\partial Q}{\partial \theta}\right)
+\omega^2 \frac{r^4}{\Delta} Q & = & 0
~~~.
\label{eq1}
\eeqa

With the ansatz

\beqa
Q(t,\theta ) & = & Q(r) C_{l+2}^{-\frac{3}{2}} (\theta )
~~~,
\label{eq3}
\eeqa
where $C_{l+2}^{-\frac{3}{2}}$ is a Gegenbauer function, we 
obtain for the final equation 

\beqa
\Delta \frac{d}{dr}\left( \frac{\Delta }{r^4} \frac{d Q}{dr}\right) - \mu^2 \frac{\Delta}{r^4} Q + \omega^2 Q & = & 0
\nonumber \\
{\rm with} &&
\nonumber \\
\mu^2 ~=~ 2n ~=~ (l-1)(l+2) &&
~~~.
\label{final-}
\eeqa
Further, setting 

\beqa
Q(r) & = & r Z^{(-)}
\label{eq4}
\eeqa
and defining ($r_*$ is the Tortoise coordinate)

\beqa
\frac{d}{d r_*} & = & \frac{\Delta}{r^2} \frac{d}{dr}
~~~,
\label{tortoise}
\eeqa
we arrive finally at the Regge-Wheeler equation \cite{RW}

\beqa
\left(\frac{d^2}{d r_*^ 2} + \omega^2 \right) Z^{(-)} & = & V^{(-)} Z^{(-)}
~~~.
\label{zerilli}
\eeqa
The Eq. (\ref{tortoise}) is the 
definition of the {\rm Tortoise coordinate}, which for an $r$
dependent $m$ has not the simple form as the one 
exposed in the book of Chandrasekhar \cite{chandra}
and in \cite{chandra1975a}. Later it will
be shown that in certain cases also analytic 
solutions may exist.

The potential $V^{(-)}$ is derived in the Appendix A
and is given by

\beqa
V^{(-)}(r) & = & 
\mu^2 \frac{\Delta}{r^4} - \frac{\Delta}{r} \frac{d}{dr}\left( \frac{\Delta}{r^4}\right)
~~~.
\label{pot-}
\eeqa

Using the definition of $\Delta$ and applying the derivatives leads to

\beqa
V^{(-)}(r) & = & \frac{\Delta}{r^5} \left[ (\mu^2 +2)r -6 m(r)+2 m^\prime (r) r \right]
~~~.
\label{pot-2}
\eeqa
The prime indicates a derivation in $r$ and (\ref{pot-2}) reduces to the result by Chandrasekhar 
\cite{chandra,chandra1975a} when
the derivative of $m(r)$ is zero, i.e., when it is constant.
The upper index $(-)$ refers to axial (negative parity)
modes.

Thus, the derivation of the equation for the axial modes is in complete analogy to the derivation presented
in \cite{chandra}. The changes in the formulas are minimal.

\section{Polar modes: Zerilli equation}
\label{polarmodes}

In contrast to the axial modes, 
for obtaining the differential equation of the polar modes the procedure is more involved.
The final equation will have a similar form as in 
(\ref{zerilli}), however, with a quite complex potential. 
S. Chandrasekhar proved \cite{chandra} that the frequencies of the polar oscillations are the same as for axial modes, which is one
of the reasons in most cases only the axial modes are 
calculated. 
Using the extended mass-function, It will be shown that
axial and polar modes are not equal anymore, though,
some similarities can be conjectured. For that reason, the
axial and polar modes have to be treated separately.
Again, we will closely follow the path exposed in 
\cite{chandra} and mainly
mention key points, deviations and approximations.

To obtain the Zerilli equation \cite{zerilli} 
for the polar modes, we proceed in the same manner as in 
\cite{chandra}, section
24b and of \cite{chandra1975a}. 
The variations 
$\delta R^{0}_{~2}$, $\delta R^{0}_{~3}$, 
$\delta R^{2}_{~3}$, $\delta G^{2}_{~2}$ and 
$\delta R^{1}_{~1}$ lead to the identical
equations as given in \cite{chandra}
($\delta R^\mu_{~\nu} = 0$, $\delta G^\mu_{~\nu}=0$ and
see discussion in Section \ref{schwarz}). New functions are introduced, varying $\nu$, $\mu_2$, $\mu_3$ and $\psi$
(equations (36)-(39) in chapter 24 of \cite{chandra}),
restricting to the quadrupole mode ($l=2$) of the
multipole expansion:

\beqa
\delta \nu & = & N(r) P_l({\rm cos}\theta ) 
\nonumber \\
\delta \mu_2 & = & L(r) P_l({\rm cos}\theta )
\nonumber \\
\delta \mu_3 & = & \left[ T(r) P_l + V(r) P_{l\mid\theta\mid\theta}\right]
\nonumber \\
\delta\psi & = & \left[ T(r) P_l + V(r) P_{l\mid \theta}{\rm cot}\theta \right]
~~~.
\label{eq2.1}
\eeqa
We are also lead to the relation (equation (43) in 
\cite{chandra})

\beqa
T-V+L & = & 0
~~~,
\label{eq2.2}
\eeqa
which reduces the number of linear independent functions
by one.  

The following steps in \cite{chandra} will remain the same. One reason is that the general structure is not affected by the modified metric term $e^{2\nu}$, 
e.g., only the function $\nu$ and its derivative $\nu_{\mid r
}=\nu^\prime$ appear and not their explicit 
dependence on $m(r)$,
it is {\it implicit}. In Eq. (48) of \cite{chandra} a new function is defined in substitution of $V$, namely

\beqa
X & = & n V ~=~ \frac{1}{2}(l-1)(l+2)V
~~~.
\label{eq2.3}
\eeqa

A relation of the derivatives of these defined functions is given in (52)-(54) of \cite{chandra}, which we will repeat here, because the coefficients in these equations do depend on $m(r)$ and, which is new, its radial derivative $m^\prime (r)$:

\beqa
N_{\mid r} & = & a N + b L +c X
\nonumber \\
L_{\mid r} & = & \left( a- \frac{1}{r} + \nu_{\mid r}\right) N + \left(b-\frac{1}{r} - \nu_{\mid r}\right) L +c X
\nonumber \\
X_{\mid r} & = & -\left(a-\frac{1}{r}+\nu_{\mid r} \right) N - \left( b + \frac{1}{r} - 2\nu_{\mid r} \right) L 
-\left( c + \frac{1}{r} - \nu_{\mid r}\right) X
~~~.
\nonumber \\
\label{eq2.4}
\eeqa
The coefficients and $\nu_{\mid r}$,
have now new contributions due to the mass-function,
and are given by

\beqa
a & = & \frac{n+1}{r-2m(r)}
\nonumber \\
b & = & -\frac{1}{r} - \frac{n}{r-2m(r)} + \frac{m(r)}{r(r-2m(r))} + \frac{m^2(r)}{r(r-2m(r))^2}
+ \omega^2 \frac{r^3}{(r-2m(r))^2}
\nonumber \\
&& - \frac{m^\prime (r)}{r-2 m(r)} - \frac{2m(r)m^\prime (r)}{(r-2m(r))^2} + \frac{r( m^\prime (r))^2}{(r-2m(r))^2}
\nonumber \\
c & = & -\frac{1}{r} + \frac{1}{r-2m(r)} + \frac{m^2(r)}{r(r-2m(r))^2} + \omega^2 \frac{r^3}{(r-2m(r))^2}
\nonumber \\
&& +\frac{m^\prime (r)\left[ -2 m(r) + m^\prime(r) r\right]}{(r-2m(r))^2} 
\nonumber \\
\nu_{\mid r} & = & \frac{m(r)}{r(r-2m(r))}
\nonumber \\
&&- \frac{m^\prime (r)}{r-2 m(r)}
~~~.
\label{eq2.5}
\eeqa
In the second row of each factor the new contributions appear, if any, depending on the derivatives of $m(r)$.
{\it The results
reduce to the one in} \cite{chandra} 
{\it when $m^\prime (r)$ is set to zero.}

In what follows, we write some of the expressions needed to derive the Zerilli equations in explicit form, 
because of
new contributions due to the dependence of $m(r)$ on $r$:

\beqa
\left(L+X\right)_{\mid r} & = & -\left( \frac{2}{r} - \frac{m(r)}{r(r-2m(r))} + \frac{m^\prime (r)}{r-2m(r)}\right) L
\nonumber \\
&& -\left( \frac{1}{r} - \frac{m(r)}{r(r-2m(r))} + \frac{m^\prime (r)}{r-2m(r)}\right) X
\nonumber \\
& = & 
 -\left( \frac{2}{r} - \frac{m(r)}{r(r-2m(r))} + \frac{m^\prime (r)}{r-2m(r)}\right) \left( L+X\right)
+ \frac{n}{r}V
\nonumber \\
&=& -\frac{1}{r(r-2m(r))} \left\{
\left[ 2r-5m(r)+rm^\prime (r)\right] L
\right.
\nonumber \\
&& \left.
+\left[ r-3m(r)+r m^\prime(r)\right] X
\right\}
\nonumber \\
X_{\mid r} & = & -\frac{\left[nr + 3m(r)-r m^\prime (r)\right]}{r(r-2m(r))} N - \frac{(n+1)}{r-2m(r)} X
\nonumber \\
&& -\left[ - \frac{m(r)}{r(r-2m(r))}+ \frac{\left(m^2(r)+\omega^2 r^4\right)}{r(r-2m(r))^2}
+ \frac{m^\prime (r)}{r-2m(r)} 
\right.
\nonumber \\
&&\left.
-\frac{2m(r)m^\prime (r)}{(r-2m(r))^2} + \frac{r (m^\prime (r))^2}{(r-2m(r))^2} - \frac{n}{r-2m(r)}
\right] \left( L + X \right)
\nonumber \\
\label{eq2.6}
\eeqa

In order to obtain the Zerilli equation in GR, one defines a function $Z^{(+)}$ as 
a particular combination of 
$N$, $V$, $X$ and $L$ (see (58) and (59) in \cite{chandra}).
After that one calculates the first and second derivative with respect to $r_*$, using the above equations of derivatives 
for $N$, $L$ and $X$. 

In \cite{chandra,chandra1975a} only the ansatz for $Z^{(+)}$
is presented, without a derivation. Here, we provide the
foundation for this ansatz of the general Zerilli equation,
which includes in the limit of a constant mass-function
the GR.

The combination of $Z^{(+)}$ in terms of the functions
$V$, $L$ and $X$ is chosen such that, when the second order
derivative in $r_*$ is applied,
the only contributions left is solely proportional
to $Z^{(+)}$. Due to the new contributions in the derivatives of $m(r)$, the ansatz for the
linear combination changes to

\beqa
Z^{(+)} & = & \alpha (r) N +\beta (r) V + \gamma (r) (L+X)
~~~.
\label{eq2.7}
\eeqa
 
On how the functions $\alpha (r)$, $\beta (r)$ and
$\gamma (r)$ are determined is explained in the
Appendix B.
Also, the ansatz proposed by S. Chanbrasekhar
\cite{chandra} will be derived,
 in the limit of $m(r)=m_0$.

In a first step, the first derivative 
($\frac{dZ^{(+)}}{dr_*}$) and 
second derivative ($\frac{d^2Z^{(+)}}{dr_*^2}$) have to be calculated,
using (\ref{tortoise}). The factors proportional to
$\omega^2$ are determined and a solution of
$\alpha (r)$ and $\beta (r)$ is found. In a second
step a differential equation for $\gamma (r)$ is set up,
where the solution will depend on the  mass-function
used. For a constant mass the GR solution of 
\cite{chandra,chandra1975a} is recovered.
 This is a rather lengthy, but straightforward, calculation done in the Appendix 
B and C.
but better done with MATHEMATICA \cite{mat11,matdetails}.

In the first step, for $\alpha (r)$ and
$\beta (r)$ we obtain (see Appendix B)

\beqa
\alpha (r) & = & 0 
\nonumber \\
\beta (r) & = & r
~~~.
\label{albetga}
\eeqa

In the second step, we take the $\omega^2$ 
{\it independent term}, obtained after
having applied
$\left[ \frac{d^2}{dr_*^2} + \omega^2 \right]$ to
$Z^{(+)}$, which leads to the 

\beqa
V_N^{(+)} N(r)+V^{(+)}_V r V(r) + V^{(+)}_{LX} \gamma (r) (L(r)+X(r)) 
~~~,
\label{ap1}
\eeqa
where $V_N^{(+)}$ depends on $\alpha (r)=0$, 
$\beta (r)=r$ and $\gamma (r)$.

Because $\alpha (r)=0$, the $Z^{(+)}$ is only a combination
in $V(r)$ and $(L(r)+X(r))$, the factor of
$N(r)$ has to vanish. This condition leads to

\beqa
\gamma (r) & = & 
-\frac{r^2}{\left(n r + 3m(r) -r m^\prime (r)\right)}
\left( 1 + \frac{2 m^\prime (r) - r m^{\prime\prime}(r)}{n}
\right)
~~~.
\label{gam-r}
\eeqa

For the linear combination in $V(r)$ and $(L(r)+X(r))$ to
be written as $V^{(+)}Z^{(+)}$, the two potential factors
in (\ref{ap1}) have to be equal, which leads to the condition

\beqa
G(r) & = & \frac{1}{2r^5 \gamma (r)}
\left\{ (-12 \gamma(r)^2 (r - 2 m(r)) (-3 m(r) + r (2 + m^\prime (r)))
\right. 
\nonumber \\
&& \left.
+ 2 r \gamma (r) (r - 2 m(r)) (m(r) (23 - 8 m^\prime (r))
\right.
\nonumber \\
&&	\left.			
+ r (-6 + 4 \gamma^\prime (r) - 11 m^\prime (r) +
        r m^{\prime\prime}(r)) 
+ r^2 (8 m(r)^2 (6 \gamma^\prime (r) - 
        r \gamma^{\prime\prime} (r))
				\right.
\nonumber \\
&& \left.
+ 4 r m(r) (-1 + 2 m^\prime (r) - 
        2 \gamma^\prime (r) (5 + 2 m^\prime (r)) 
\right.
\nonumber \\
&& \left.
+ 2 r \gamma^{\prime\prime}(r) - r m^{\prime\prime}(r)) + 
     r^2 (2 (1 + 4 \gamma^\prime (r) - 
		2 m^\prime (r)) (1 +
            m^\prime (r)) 
\right.
\nonumber \\
&& \left.
- 2 r \gamma^{\prime\prime}(r) + 
        r (1 + 2 m^\prime (r)) m^{\prime\prime}(r)))
\right\}
= 0 ~~~.
\label{dif-gam}
\eeqa
As noted above, for a constant mass $m(r)=m_0$, this
equation is identically fulfilled for the expression{
given in \cite{chandra,chandra1975a}.
In Appendix C we will see
that this solution satisfies the condition (\ref{dif-gam})
for a wide range of $r$, save
near $r=\frac{3}{2}m_0$, with a small error, though. 
This shows that the Zerilli equation can be constructed
approximately. 

Finally, the Zerilli equation acquires the form
 
\beqa
\left[ \frac{d^2}{dr_*^2} + \omega^2 \right] Z^{(+)} & = & V^{(+)} Z^{(+)}
~~~,
\label{zerilli+}
\eeqa
which describes the polar modes with an $r$ dependent 
mass-function $m(r)$. As we will see 
further below, the axial and polar modes, though different, 
still share similar structures.

In terms of the mass-function $m(r)$, using (\ref{gam-r})
the potential for the polar modes is given by

\beqa
&
V^{(+)}(r) =
&
\nonumber \\
&
((r - 2 m (r )) (18 (m (r))^3 + 
          3 r (m (r) )^2 (6 n - 12 m^\prime (r ) + 
               5 r m^{\prime\prime} (r ) - 
&
\nonumber \\
&  
               4 r^2 m^{\prime\prime\prime} (r )) + 
     r^3 (2 n^2 (1 + n) - 
        8 (m^\prime (r ))^3 + (m^\prime (r ))^2 (8 + 6 n + 
           3 r m^{\prime\prime} (r )) + 
&
\nonumber \\
& 
        y (m^{\prime\prime} (r ) (-n (6 + n) + 
              2 r m^{\prime\prime} (r )) + 
           2 n r m^{\prime\prime\prime} (r )) 
&
\nonumber \\
& 
        -2 m^\prime (r ) (-4 n + 
           r ((3 + n) m^{\prime\prime} (r ) + 
              r M^{\prime\prime\prime} (r )))) + 
&
\nonumber \\
& 
     2 r^2 m (r ) (3 n^2 + 9 (m^\prime (r))^2 + 
        r (m^{\prime\prime} (r ) (-3 + 5 n - 
              2 r m^{\prime\prime} (r ))  
&
\nonumber \\
& 
+ (3 - 2 n) r m^{\prime\prime\prime} (r )) +
				m^\prime (r ) (-12 n + 
             r (m^{\prime\prime} (r ) + 
                2 r m^{\prime\prime\prime} (r ))))))/
&
\nonumber \\								
&		
(r^4 (3 m(r) + r (n - m^{\prime} (r )))^2)
~~~.
&
\label{pot+}
\eeqa

In what follows, the dimensionless coordinate

\beqa
y & = & \frac{r}{m_0}~=~\frac{y_{eh}}{1-\xi}
\label{yeh}
\eeqa
is used,
where $y_{eh}$ is the position of the event horizon
and the
variable $\xi$ has the range $[0,1]$. When $\xi=0$, then 
$y=y_{eh}$
and when $\xi$ tends to 1, the coordinate $y$ 
tends to $+\infty$. The particular mass-function used,
corresponds to an event horizon at 
$y_{\rm eh}=\frac{3}{2}$.

The reason for using the coordinate $\xi$ with a compact
support lies in the use of the AIM method, explained further
below. It guarantees a better convergence of an iterative
equation.

\section{Constructing 
the Regge-Wheeler and Zerilli equations
}
\label{zerr}

In this section the final form of the differential equations, 
used for the axial and polar modes, will
be derived.

In subsection 5.2
the {\it Asymptotic Iteration Method} (AIM)
is resumed, which solves
a differential equation of second order. 
In what follows, 
the Regge-Wheeler/Zerilli equation is rewritten in
a form, which is practical for the AIM. 

For the function $m(r)$
the particular form, defining $y=\frac{r}{m_0}$,

\beqa
m(r) & = & m_0m(y)
\nonumber \\
m(y) & = & 1- \frac{27}{32 y^{3}}
~~~,
\label{eq11}
\eeqa
is used, which exhibits an event horizon at $y=\frac{3}{2}$.

\subsection{Rewriting the differential equation}

First, the explicit form of the differential equation is derived, noting that the Regge-Wheeler and
Zerilli equation can be, in general, written as

\beqa
\left[ \frac{d^2}{dr_*^2} + \omega^2 - 
V^{(\pm )}(r)\right] Z^{(\pm )} & = & 0
~~~.
\label{aim1}
\eeqa

Let us use, for a moment, the more general mass-function
$m(y)=\left(1-\frac{b}{12 y^3}\right)$, which for
$b=\frac{81}{8}$ acquires the one of (\ref{eq11}).
It was used in \cite{universe} for the study of 
phase transitions from GR to pc-GR.
The relation of $y=\frac{r}{m_0}$ to a variable
$\xi$ with a compact support for the range of integration
is defined as in (\ref{yeh}),
where $y_{eh}(b)$ is the position 
of the event horizon as a function of the parameter $b$. 
It is the solution of the condition
for the event horizon in the Schwarzschild case, i.e., 

\beqa
y^4 - 2 y^3 + \frac{b}{6} & = & 0
~~~.
\label{AIM3}
\eeqa  

Using the Wolfram MATHEMATICA code 
\cite{mat11,matdetails}, the solution is 

\beqa
y_{eh}( b ) & = & 
\frac{1}{2} + \frac{1}{2\sqrt{3}}\left[\left(
 3 + \frac{2 b}{\left(9 b + \sqrt{81 b^2 - 8 b^3}\right)^{\frac{1}{3}}} 
+ \left(9 b + \sqrt{81 b^2 - 8 b^3}\right)^{\frac{1}{3}}
\right)\right]^{\frac{1}{2}}
\nonumber \\
&& + \frac{1}{2} \left[
\left(
2 - \frac{2 b}{3 \left(9 b + \sqrt{81 b^2 - 8 b^3}\right)^{\frac{1}{3}}}
     -\frac{1}{3} \left(9 b + \sqrt{81 b^2 - 8 b^3}\right)^{\frac{1}{3}}
\right.\right.
\nonumber \\
& &
\left.\left. 
+\frac{2 \sqrt{3}}
{
\left(
\left[\left(3 + \frac{2 b}{\left(9 b + \sqrt{81 b^2 - 8 
b^3}\right)^{\frac{1}{3}}} 
+\left(9 b + \sqrt{81 b^2 - 8 b^3}\right)^{\frac{1}{3}}
\right)
\right]^{\frac{1}{2}} \right)
} \right)\right]^{\frac{1}{2}}
~~~.
\label{AIM4}
\eeqa

The second order derivative with respect to the Tortoise coordinate is 

\beqa
m_0^2\frac{d^2}{dr_*^2} & = &
\frac{\Delta}{r^2}\frac{d}{dr}\frac{\Delta}{r^2}\frac{d}{dr}
\nonumber \\
& = & 
\left(1-\frac{2}{y}m(y)\right)^2 \frac{d^2}{dy^2} 
+ \left(1-\frac{2}{y}m(y)\right)\frac{2}{y^2}\left(m (y) - y m^\prime (y)\right) \frac{d}{dy}
~~~.
\nonumber \\
\label{AIM5}
\eeqa
The prime refers to a derivative in $y$.

Defining the dimensionless expressions 

\beqa
{\tilde \omega} & = & m_0 \omega ~,~ 
{\tilde V}(y) ~=~ m_0^2 V(r)
~~~,
\label{AIM6}
\eeqa
the differential equation acquires the form

\beqa
\left[ \left( 1-\frac{2}{y} m(y)\right)^2 \frac{d^2}{dy^2} + 
\right. &&
\nonumber \\
\left.
\left(1-\frac{2}{y}m(y)\right) \frac{2}{y^2}
\left( m(y) - y m^\prime (y)\right) \frac{d}{dy} + {\tilde \omega}^2 - {\tilde V}^{(\pm )}(y)\right] Z(y) & = & 0
~~~.
\label{AIM7}
\eeqa

The next step is to substitute $y$ by $\xi$.
We also define $m(\xi )$ as $m(y(\xi ))$ and $m^\prime (\xi )$
as the same as $m^\prime (y)$
(in order to avoid more definitions of functions, we use
the same letter $m$).
Doing so, leads after some manipulations to

\beqa
\left[ \frac{d^2}{d\xi^2} -
\frac{2}{(1-\xi)}\left( 1 - \frac{(1-\xi)}{y_{eh}(b)} \frac{\left( m(\xi ) - (y_{eh}/(1-\xi )) m^\prime (\xi)\right)}
{\left(1- \frac{2(1-\xi )}{y_{eh}(b)} m(\xi )\right)} \right) \frac{d}{d\xi}
\right. &&
\nonumber \\
\left.
\frac{(y_{eh}(b))^2}{(1-\xi )^4} 
\frac{\left( {\tilde \omega}^2 
- {\tilde V}^{(\pm )}(\xi )\right)}{\left( 1 
- \frac{2(1-\xi )}{y_{eh}(b)} m(\xi ) \right)^2}
\right] Z( \xi ) & = & 0
~~~.
\nonumber \\
\label{AIM8}
\eeqa
Remember that the prime refers to the derivative with respect to $y$ and {\it not}to  $\xi$.)
Note also, that
$V^{(\pm )}(r) = V^{(\pm )}(y(\xi ))/m_0^2$,
where $V^{(\pm )}(y(\xi ))$ is dimensionless.

In practical calculations the function 
(\ref{eq11}) is used, which results in
the function $m(\xi )$

\beqa
m(\xi ) & = & 1 - \frac{(1-\xi )^{3}}{4}
~~~.
\label{AIM9}
\eeqa

For the potentials $V^{(\pm )}$ in terms of $\xi$, we obtain,
the expressions enlisted in \cite{matdetails}.

\subsection{AIM}
\label{AIM}

The AIM was introduced by H. Ciftci, R. L. Hall, and N. Saad 
 \cite{ciftci2003,ciftci2005}, for solving
second order differential equations of the form \cite{cho2012}

\beqa
f^{\prime\prime}(\xi ) & = & \lambda_0 (\xi ) f^\prime (\xi )
+ s_0(\xi ) f(\xi ) 
~~~,
\label{eq10}
\eeqa
with $\xi$ as the variable.

Deriving both sides $p$ times, $p$ being an integer, 
leads to an equivalent differential equation

\beqa
f^{(p+1)}(\xi ) & = & \lambda_{p-1} (\xi ) f^\prime (\xi )
+ s_{p-1}(\xi) f(\xi ) 
~~~,
\label{eq10a}
\eeqa
with

\beqa
\lambda_p(\xi ) & = & \lambda^\prime_{p-1}(\xi ) 
+ s_{p-1}(\xi )
+\lambda_0(\xi )\lambda_{p-1}(\xi )
\nonumber \\
s_p(\xi ) & = & s^\prime_{p-1}(\xi ) 
+ s_0(\xi )\lambda_{p-1}(\xi )
~~~.
\label{eq12}
\eeqa
Convergence is achieved, when the ratio of $s_p(x)$ and 
$\lambda_p(x)$ does not change, with $p$ the
iteration number. Once achieved, the 
{\it Quantization Condition} reads

\beqa
s_p(\xi )\lambda_{p-1}(\xi )-s_{p-1}(\xi )\lambda_p(\xi ) 
& = & 0
~~~.
\label{eq13}
\eeqa
This expression depends on $\xi$, which can be chosen arbitrarily,
and has to be resolved for the frequencies $\omega$. 
Because of its $\xi$-dependence, it is a very subtle task
to obtain convergence rapidly. The problem was resolved partially in \cite{cho2012}, 
expanding the $\lambda_p$ and
$s_p$ in a Taylor series around a point $\xi$, defining

\beqa
\lambda_p(\rho ) & = & \sum_{i=0}^\infty c_p^i (\xi -\rho )^i
\nonumber \\
s_p(\rho ) & = & \sum_{i=0}^\infty d_p^i (\xi -\rho )^i
~~~.
\label{eq14}
\eeqa

Substituting this into (\ref{eq12}) leads to a new recursion
relation for the coefficients $c_p^i$ and $d_p^i$ and a new 
quantization condition

\beqa
d_p^0 c_{p-1}^0 - d_{p-1}^0 c_p^0 & = & 0
~~~.
\label{eq15}
\eeqa

The clear advantage of (\ref{eq15}) lies in the fact that
(\ref{eq15})
depends only on the frequencies and is a polynomial
in the frequencies. Thus, the determination of the 
$\omega$-spectrum is restricted to solve (\ref{eq15}). 
However, the result 
still depends
on the point of expansion $\xi$ 
in the Taylor series (\ref{eq14}).

In order to obtain a "quick" convergence, the following rules should be observed, which are the results of the experience 
of others
\cite{cho2012} and ours:

\begin{itemize}

\item A compact support for the range of the coordinate should be used, 
i.e. the coordinate $\xi$, which is zero at the event horizon and
approaches 1 for $r\rightarrow\infty$. 
In a Taylor expansion, this prevents too
large deviations to the real potential function at the limits
$\xi =0$ and $\xi =1$. I.e., it is important to describe
the potential well near the limits.

\item The asymptotic behavior for 
$r_* \rightarrow \pm\infty$ 
of the wave function should be extracted as exactly as possible. Analytic solutions are of course the best.

\item For the expansion around a point $\xi$, the maximum or minimum of the potential
is recommended as a starting point. However, 
shifting it to its vicinity at 
larger values may give better convergence.
The following criterion helps, namely that with an
increasing number of iterations 
the lower frequencies are not changing
any more for low values of $-\omega_I$.

\item Using MATHEMATICA, only rational numbers 
are allowed. 
In case of irrational numbers, it is recommended to 
approximate them by rational ones, otherwise MATHEMATICA
develops numerical instabilities. 

\end{itemize}

\section{Spectrum of the Regge-Wheeler equation (axial
modes) and the Zerilli equations (polar modes)}
\label{ZE-RAIM}

The Regge-Wheeler equation
for the {\it axial modes} and Zerilli equations 
for the {\it polar modes} are solved, with the
help of the AIM. 
In a first step the asymptotic limit is discussed
and taken into account in the definition of the 
$Z^{(\pm )}$-functions, which leads 
to the final form of the differential equation
to solve. 

We will present calculations with different iteration
numbers, which allows to judge the convergence of
the iteration method and see 
trends for large iteration numbers.

\subsection{The asymptotic limit}

The wave solution must satisfy the condition

\beqa
\Psi & \rightarrow & \left\{
\begin{array}{c}
e^{+i\omega r_*} ~,~ {\rm for} ~r_* \rightarrow +\infty 
~(r \rightarrow \infty )\\
e^{-i\omega r_*} ~,~ {\rm for} ~r_* \rightarrow -\infty
~(r \rightarrow \frac{3}{2}m_0)
\end{array}
\right.
~~~.
\label{AS1}
\eeqa
The time dependence for both limits is $e^{-i\omega t}$.
This implies that for a complex 
$\omega = \omega_R + i\omega_I$, the time dependence has the form

\beqa
e^{-i\omega t} & = & e^{-i\omega_R t} e^{+\omega_I t}
~~~,
\label{AS1a}
\eeqa
i.e., for an exponential decreasing function the imaginary
part of the frequency ($\omega_I$) has to be {\it negative}
($\omega_I<0$), otherwise, there is no damping, i.e., no stable
mode. The Schwarzschild solution is stable under the
perturbations when no positive $\omega_I$ 
solution appears.

The integrated relation of the Tortoise 
coordinate, defined in (\ref{tortoise}), 
$y_*=\frac{r_*}{m_0}$ to the variable 
$y=\frac{r}{m_0}$ is given by

\beqa
y_* & = & \int \frac{d y}{\left(1- \frac{2m(y)}{y}\right)} 
~~~.
\label{AS1-1}
\eeqa

For $m(y)=1$, i.e. GR, the solution is well
known, namely

\beqa
y_* & = & y + 2{\rm ln}\left( \frac{y}{2} -1 \right)
~~~,
\label{AS2}
\eeqa

For $m(y)=\left( 1-\frac{27}{32 y^3} \right)$,
there is surprisingly also
a solution

\beqa
y_* & = & y+2{\rm ln}\left(\frac{y}{\frac{3}{2}} - 1\right)
\nonumber \\
&& +2{\rm ln}\left(3\right)-\frac{3}{2}+\frac{9}{(12-8 y)}- 
\frac{{\rm arctan}\left[\frac{1+2 y}{\sqrt{2}}\right]}
{4\sqrt{2}}
~~~,
\label{AS3}
\eeqa
which was obtained, using MATHEMATICA \cite{mat11}.
When some of the parameters in $m(y)$ are changed,
no solution can be found. That only this particular ansatz of
$m(r)$ provides an analytic solution for the Tortoise coordinate
and not any other with a different parametrization, is
quite a surprise which we would like to understand.

Changing to the variable $\xi = 1-\frac{3}{2y}$,
the relation of the Tortoise coordinate to $\xi$ is

\beqa
y_* & = & \frac{3}{2(1-\xi )} + 2{\rm ln}(\frac{\xi}{1-\xi})
\nonumber \\
&&
+ 2{\rm ln}(3)
-\frac{3}{2} - \frac{3(1-\xi )}{4\xi}
-\frac{\rm{arctan}\left[\frac{4-\xi}{\sqrt{2}(1-\xi )}
\right]}{4\sqrt{2}} 
~~~.
\label{ystar}
\eeqa

\subsubsection{ Limit $r_* \rightarrow +\infty$}

We again define $y= \frac{r}{m_0}$ ($y_*= \frac{r_*}{m_0}$) and substitute $y_*$ by (\ref{ystar}). We also define

\beqa
{\tilde \omega}& = & m_0 \omega
~~~.
\label{AS3-1}
\eeqa
For $y_* \rightarrow +\infty$, the $\xi$ tends to 1, i.e,
terms proportional to
$\frac{1}{(1-\xi )}$ and ${\rm ln}(1-\xi )$ dominate.
This leads to the asymptotic form 
($e^{i\omega r}=e^{i\widetilde{\omega}y_*}$)

\beqa
e^{+i\widetilde \omega y_*}
& \rightarrow &
e^{i\frac{3{\tilde \omega}}{2(1-\xi )}} 
(1-\xi )^{-2i{\tilde\omega}}
~~~. 
\label{AS4}
\eeqa
The other terms in (\ref{ystar}) can be neglected, because they
are either constant or approach a constant value in this limit.

\subsubsection{ Limit $r_* \rightarrow -\infty$}

In this case, the $y$ approaches the event horizon. 

Using the above obtained expression for $y_*$, 
we obtain the additional terms, taking into
account that terms proportional to 
$\frac{1}{\xi }$ and ${\rm ln}(\xi )$ dominate, namely 

\beqa
&
e^{i\frac{3\widetilde{\omega}(1-\xi)}{4\xi}} 
\xi^{-2 i \widetilde{\omega}}
~~~. 
&
\label{AS4-1}
\eeqa
The $e^{i\frac{3}{2}}$ and 
$e^{-i\widetilde{\omega}
\frac{{\rm arctan}\left[ \frac{(2+\xi )}
{\sqrt{2}(1-\xi )}\right]}{4\sqrt{2}}}$ contributions
are skipped because the first is a constant and the second
has the limit $e^{-i\widetilde{\omega}{\rm arctan}\sqrt{2}}$
for $\xi = 0$, thus, it does not effect the
asymptotic limit.

\subsubsection{Final ansatz of the asymptotic behavior}

Thus, extracting the complete asymptotic limits, the new ansatz for the wave function is

\beqa
Z^{(\pm )}(\xi ) & = & 
e^{i\frac{3{\tilde \omega}}{2(1-\xi )}} 
(1-\xi )^{-2i{\tilde\omega}}
e^{i\frac{3\widetilde{\omega}(1-\xi)}{4\xi}} 
\xi^{-2 i \widetilde{\omega}}
P^{(\pm )}(\xi )
~~~,
\label{AS10}
\eeqa
with the new wave-function $P^{(\pm )}(\xi )$, whose differential equations
are set up and solved with the help of the AIM 
\cite{matdetails}.

\subsection{Axial and polar potential}
\label{v-axial-polar}

In Fig. \ref{V-axial-polar} the potentials obtained for
the axial and polar modes are depicted. In the first row
the potential for the axial modes in the GR-case 
($m(r)=m_0$) is shown. Because in GR, the axial and polar modes are  
equal, it is sufficient to plot only this potential.
In the lower row, left panel, the potential $V^{(-)}$
in pc-GR
is depicted and in the right panel the potential 
$V^{(+)}$ for the polar modes in pc-GR. As
noted, all these potentials have similar characteristics,
{\it leading to the conjecture 
that the frequency spectrum of axial and polar modes
still may share some common structure}. However,
as we will see, the axial and polar modes are not isospectral,
which is a surprise, considering that in GR they are,
What the deep origin of this difference is cannot be 
eaxplained at this moment. Isopectrality is
related to supersymmetric transformations of a
Schr\"odinger type of potential \cite{cooper1995} but
with non-bound states. Isospectrality between the axial
and polar modes for a more general type of metrics 
(de Sitter and Anti-de Sitter) was proven in \cite{moulin2020}.
A similar procedure we plan to apply for the present study,
in order to see which potentials give rise to isospectrality.

\begin{figure}
\begin{center}
\rotatebox{0}{\resizebox{150pt}{200pt}{\includegraphics[width=0.23\textwidth]{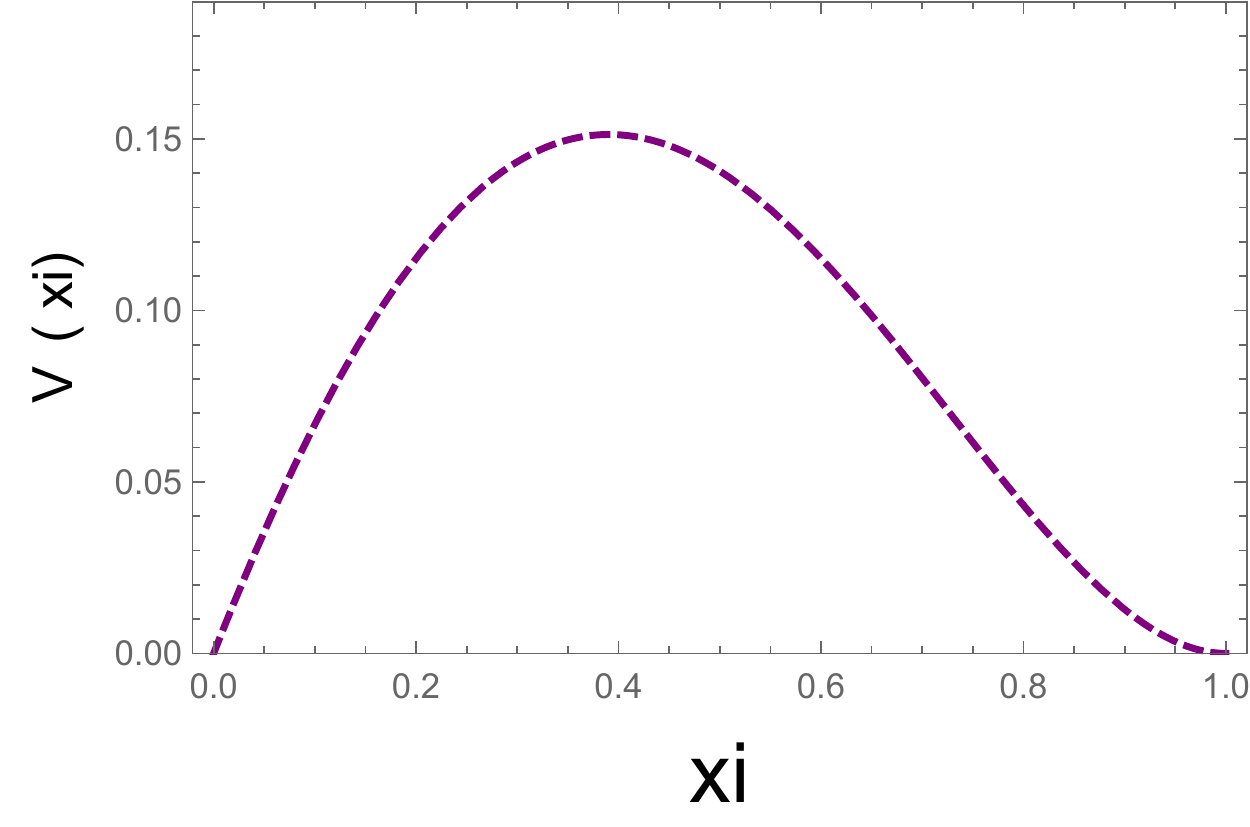}}} \\
\rotatebox{0}{\resizebox{150pt}{200pt}{\includegraphics[width=0.23\textwidth]{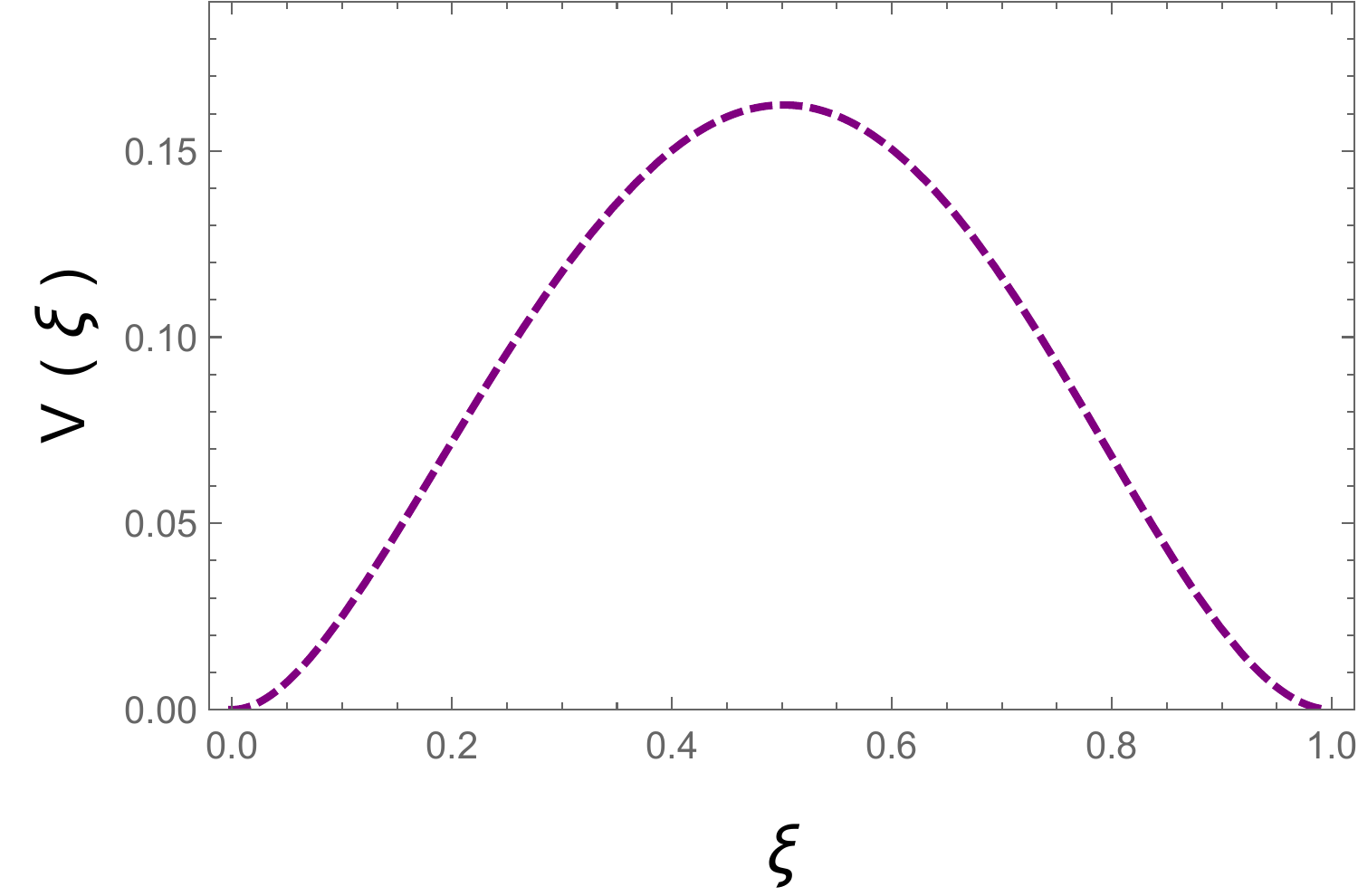}}}
\rotatebox{0}{\resizebox{150pt}{200pt}{\includegraphics[width=0.23\textwidth]{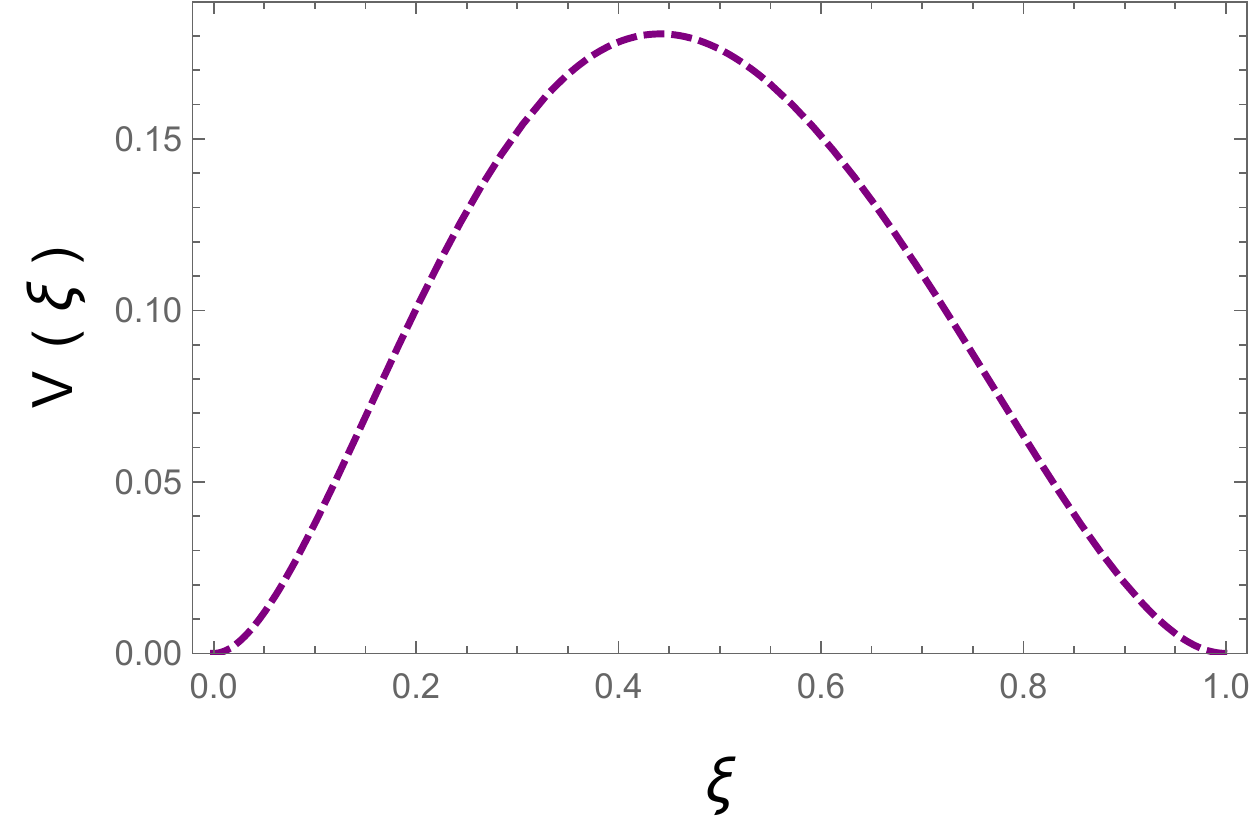}}}
\caption{
First row: Potential for the axial modes in GR.
Second row:
The potential for the axial modes ($V^{(-)}$, left panel) and 
the polar modes ($V^{(+)}$, right panel).
\label{V-axial-polar}
}
\end{center}
\end{figure}

\subsection{Spectrum of the Regge-Wheeler equation: axial modes}
\label{ZAIM-axial}

Using the AIM, we obtained the spectrum for 
198 (red dots) and 200 iterations (blue dots),
plotted in Fig. \ref{axial},
which compares GR (upper row) with pc-GR (lower row). 
While convergence is observed for
the low damping modes ($\mid \widetilde{\omega}_I\mid$ 
small), 
there is still no convergence obtained for the high damping modes. In Figure \ref{axial-400} the axial modes are depicted
for 400 iterations. Above, the GR (left panel) is compared
to pc-GR (right panel) for a large range of 
$-\widetilde{\omega}_I$. In the lower
row the same is plotted but for a restricted range of
$-\widetilde{\omega}_I$. The left panel (GR) 
reproduces the Figure 2 in \cite{kokkotash}
and of Figure 5 in \cite{konoplya2011}, where also
distinct methods to resolve the differential equation 
are resumed.

Comparing GR with pc-GR, the structure shares still some
common features, namely 
that the figure reminds at a fish with its head to the right
and its tail to the left. However, the head is moving further
to the left when the number of iterations is increased,
thus, it is not a physical property and has to be rejected.
The left panel in Fig. \ref{axial-400} shows the
result for GR and the right one for pc-GR.
Very interesting is, that the structure of a 
raising branch for pc-GR from low to large damping modes
is also stable. 
This branch can be approximated by continuum
and represents a definite feature of pc-GR,
which is not present in GR.

At low damping, convergence is obtained and the frequencies
in pc-GR are comparable in size to GR.
This changes for the large damping modes,
where the frequencies are significantly larger in pc-GR than in
GR.

\begin{figure}
\begin{center}
\rotatebox{0}{\resizebox{150pt}{200pt}{\includegraphics[width=0.23\textwidth]{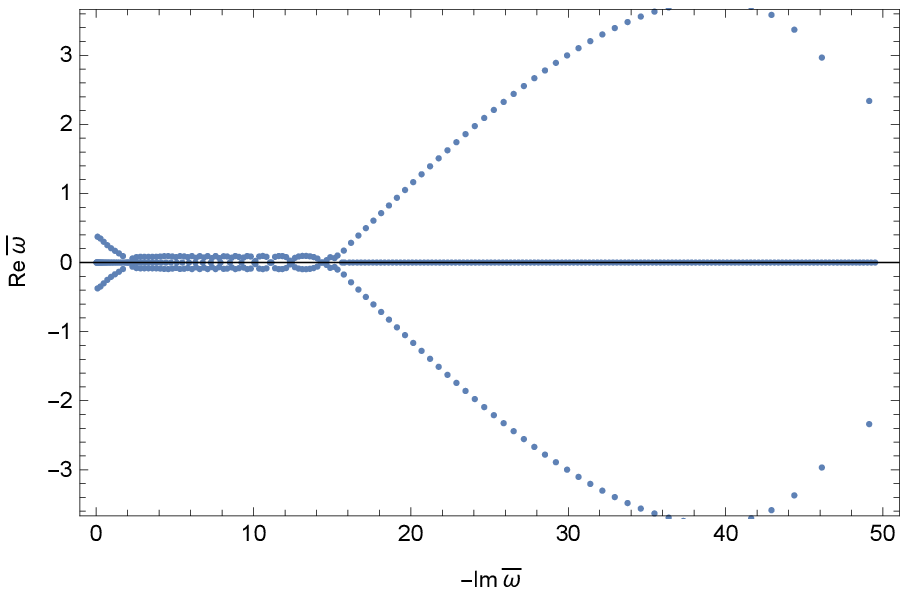}}}
\rotatebox{0}{\resizebox{150pt}{200pt}{\includegraphics[width=0.23\textwidth]{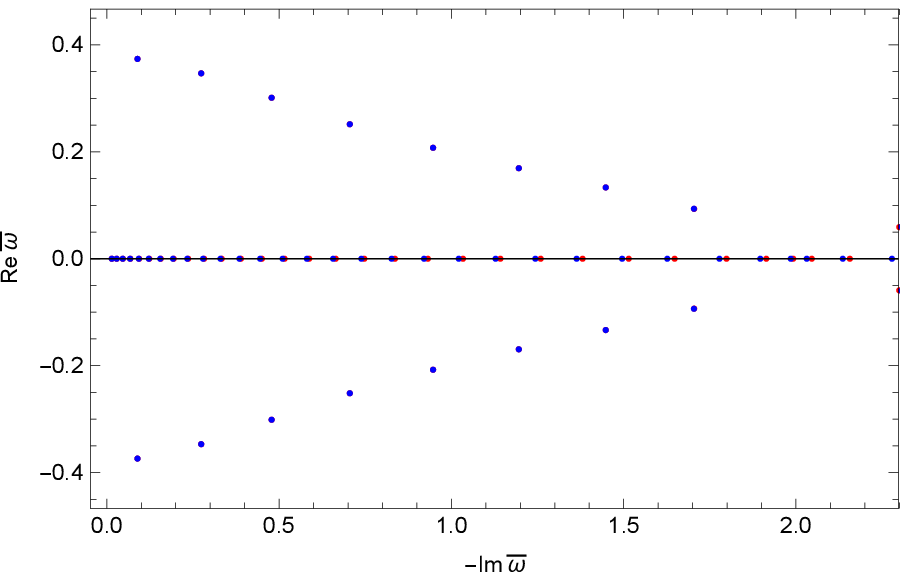}}} \\
\rotatebox{0}{\resizebox{150pt}{200pt}{\includegraphics[width=0.23\textwidth]{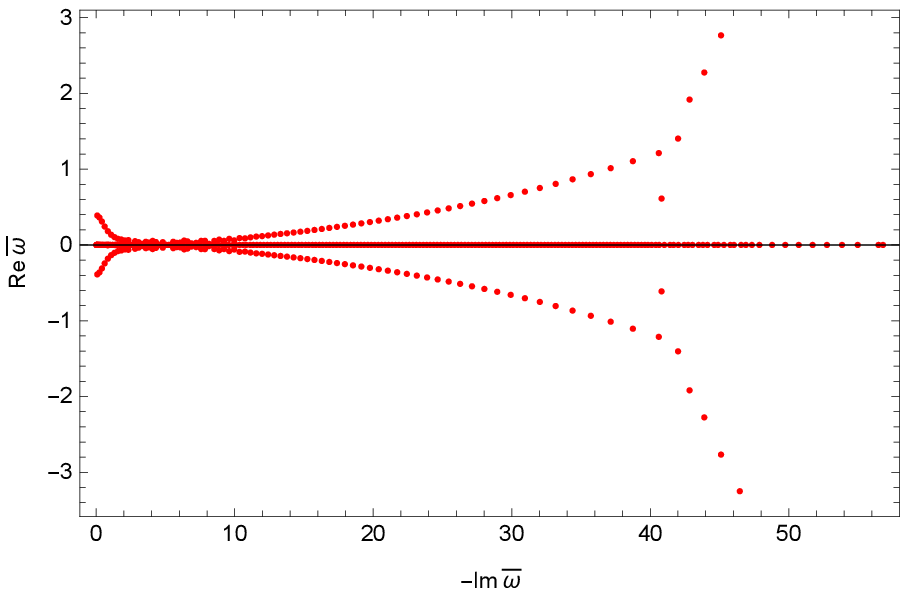}}}
\rotatebox{0}{\resizebox{150pt}{200pt}{\includegraphics[width=0.23\textwidth]{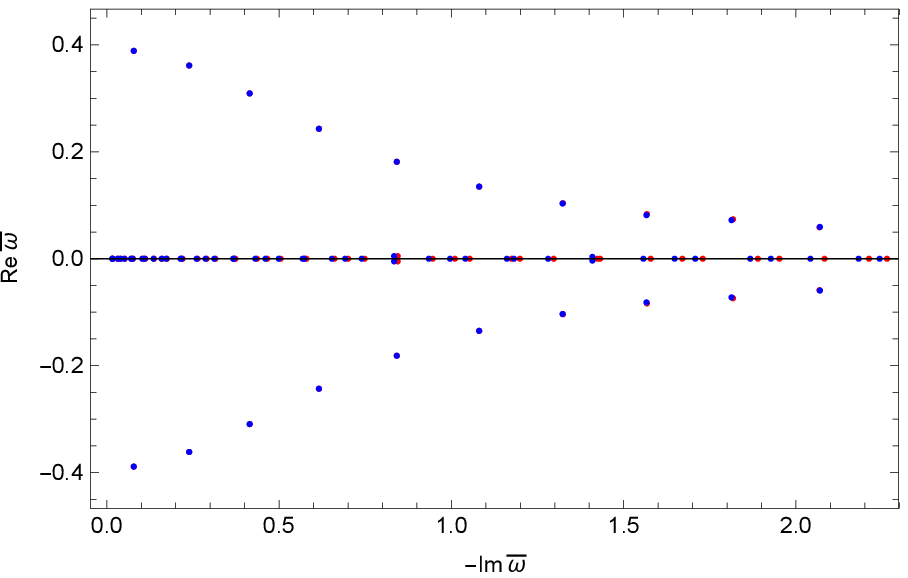}}}
\caption{
Spectrum of the axial modes for 200 iterations. 
The horizontal axis corresponds
to minus the imaginary part of the frequency, which has to
be positive in order to represent a damped, stable 
oscillation. No negative
values are present, rendering the system stable. The vertical
axis depicts the real part of the frequency.
The first row shows the axial modes in GR, while the second
row depicts the frequencies in pc-GR. 
The left panel shows the frequency distribution for a
larger range of $-\widetilde{\omega}_I$ while the
right panel is restricted to small damping modes.
\label{axial}
}
\end{center}
\end{figure}

\begin{figure}
\begin{center}
\rotatebox{0}{\resizebox{150pt}{200pt}{\includegraphics[width=0.23\textwidth]{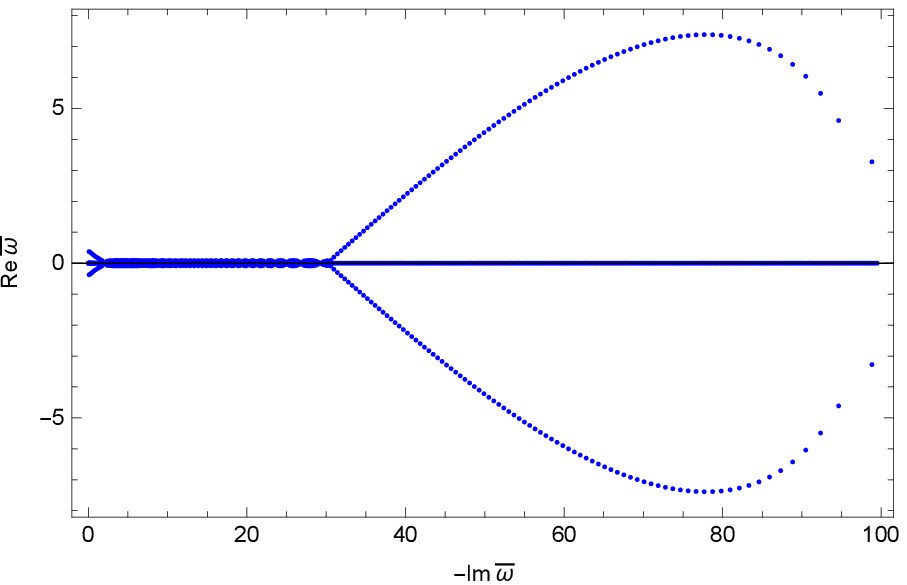}}}
\rotatebox{0}{\resizebox{150pt}{200pt}{\includegraphics[width=0.23\textwidth]{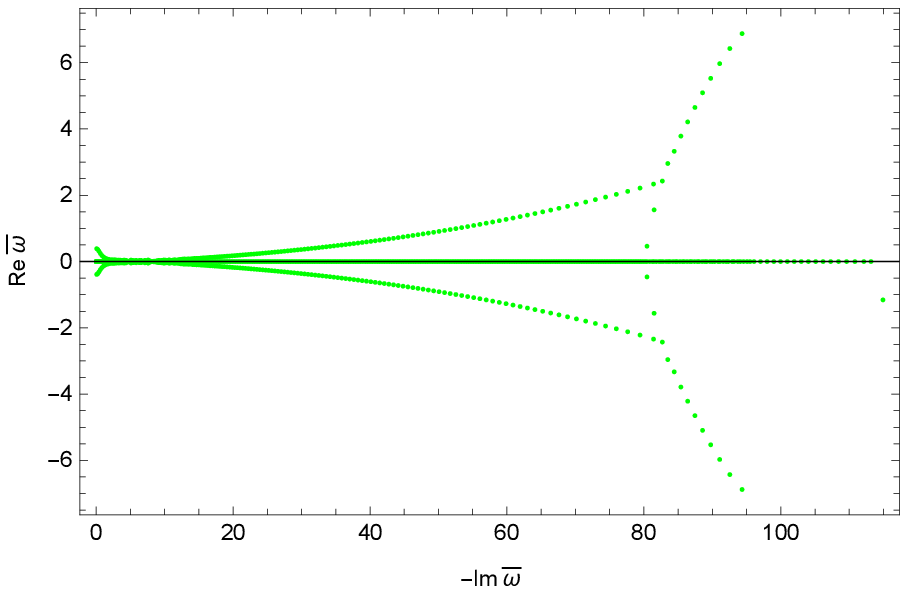}}} \\
\rotatebox{0}{\resizebox{150pt}{200pt}{\includegraphics[width=0.23\textwidth]{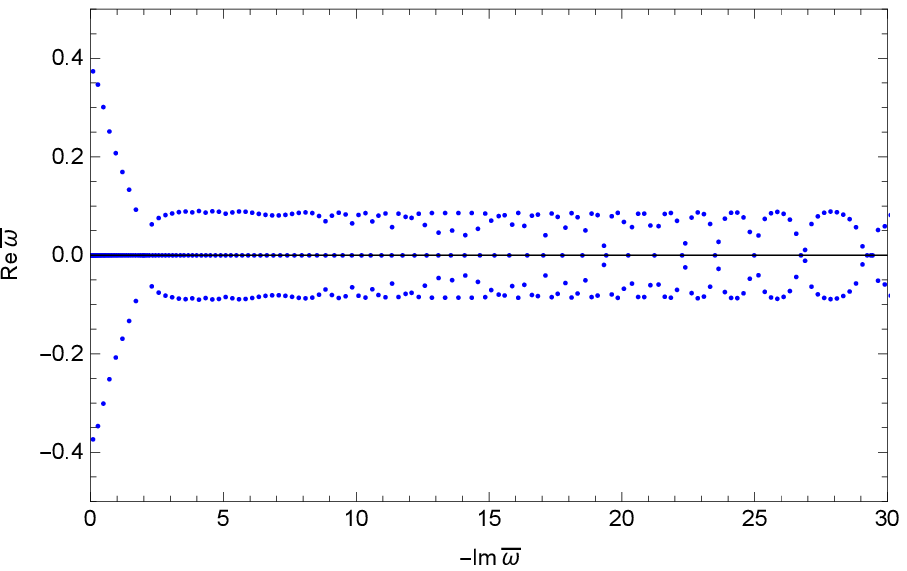}}}
\rotatebox{0}{\resizebox{150pt}{200pt}{\includegraphics[width=0.23\textwidth]{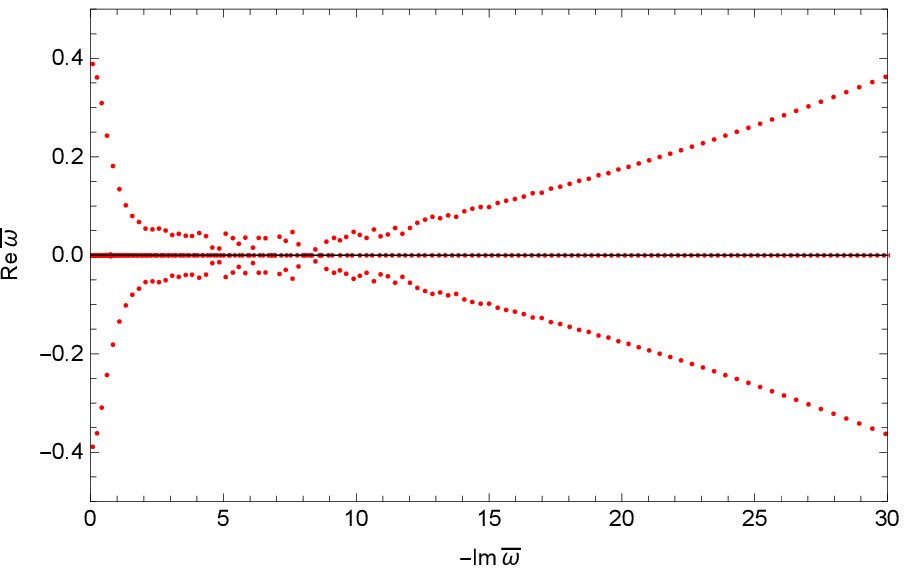}}}
\caption{
Spectrum of the axial modes for 400 iterations. 
The horizontal axis corresponds
to minus the imaginary part of the frequency and the vertical 
axis to the real part of the complex frequency. 
In each row,
the left panel is for GR and the right one for pc-GR.
Note, that compared to 200 iterations,
the onset of the "head" is shifted far to the right
in $-\widetilde{\omega}_I$, showing that this part is of no physical significance. 
For more details, in the lower row a zoom to a restricted
range is shown.
\label{axial-400}
}
\end{center}
\end{figure}

\subsection{Spectrum of the Zerilli equation: polar modes}
\label{ZAIM-polar}

In Figure \ref{polar} the polar frequency modes in pc-GR are
depicted. The structure is similar to the one for the
axial modes (see (\ref{axial-400})), 
as expected when comparing the two potentials.
In Figure
\ref{polar} the red dots correspond to 200 iterations,
the green dots to 300 and the blue dots to
400 iterations. Note, that convergence is clearly obtained
for low values of $-\widetilde{\omega}_I$ and up to 
30 the convergence is also acceptable. Thus the feature
of a raising curve for large damping is confirmed.
The "fish head", however, has moved further to the 
right, showing its unphysical nature.

\begin{figure}
\begin{center}
\rotatebox{0}{\resizebox{150pt}{200pt}{\includegraphics[width=0.23\textwidth]{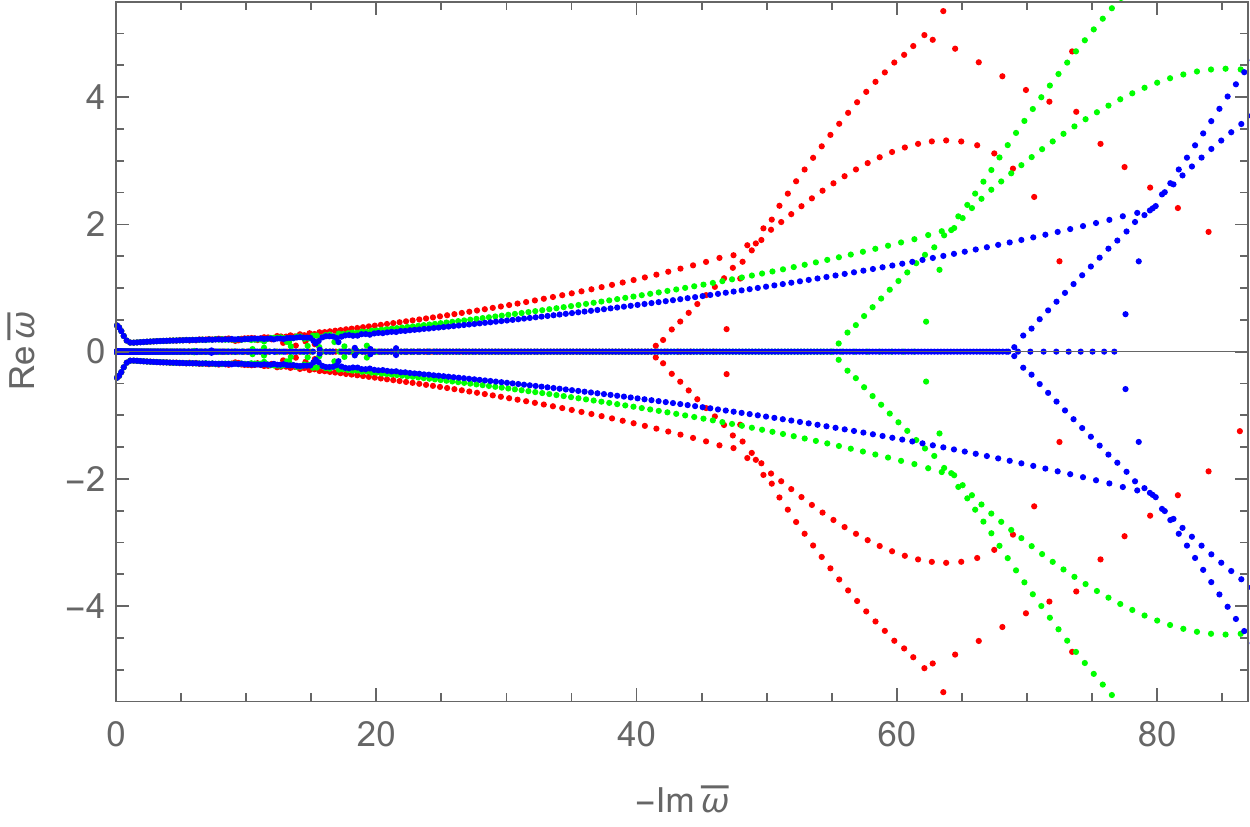}}}
\rotatebox{0}{\resizebox{150pt}{200pt}{\includegraphics[width=0.23\textwidth]{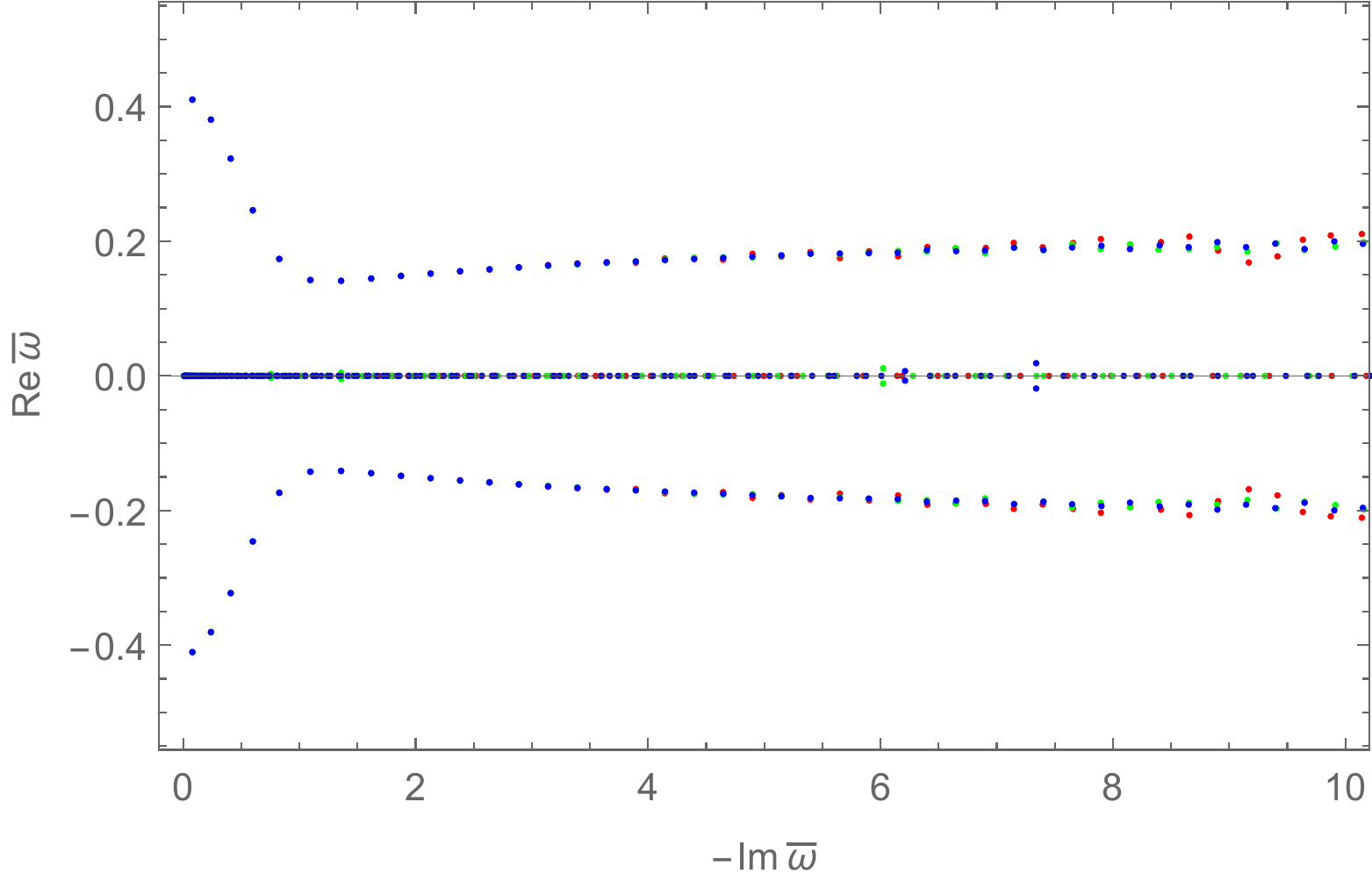}}}
\caption{
Spectrum of the polar modes for 200 (red dots), 
300 (green dots) and 400 (blue dots) iterations. 
The horizontal axis corresponds
to minus the imaginary part of the frequency, which has to
be positive for a damped oscillation. No negative
values are present, rendering the system stable. The vertical
axis depicts the real part of the frequency. 
The left panel shows the frequency distribution for a
larger range of $-\widetilde{\omega}_I$ while the
right panel is restricted to small damping modes.
Note, how the "fish-head" moves further to the right
when the iteration number is increased.
\label{polar}
}
\end{center}
\end{figure}

\begin{figure}
\begin{center}
\rotatebox{0}{\resizebox{150pt}{200pt}{\includegraphics[width=0.23\textwidth]{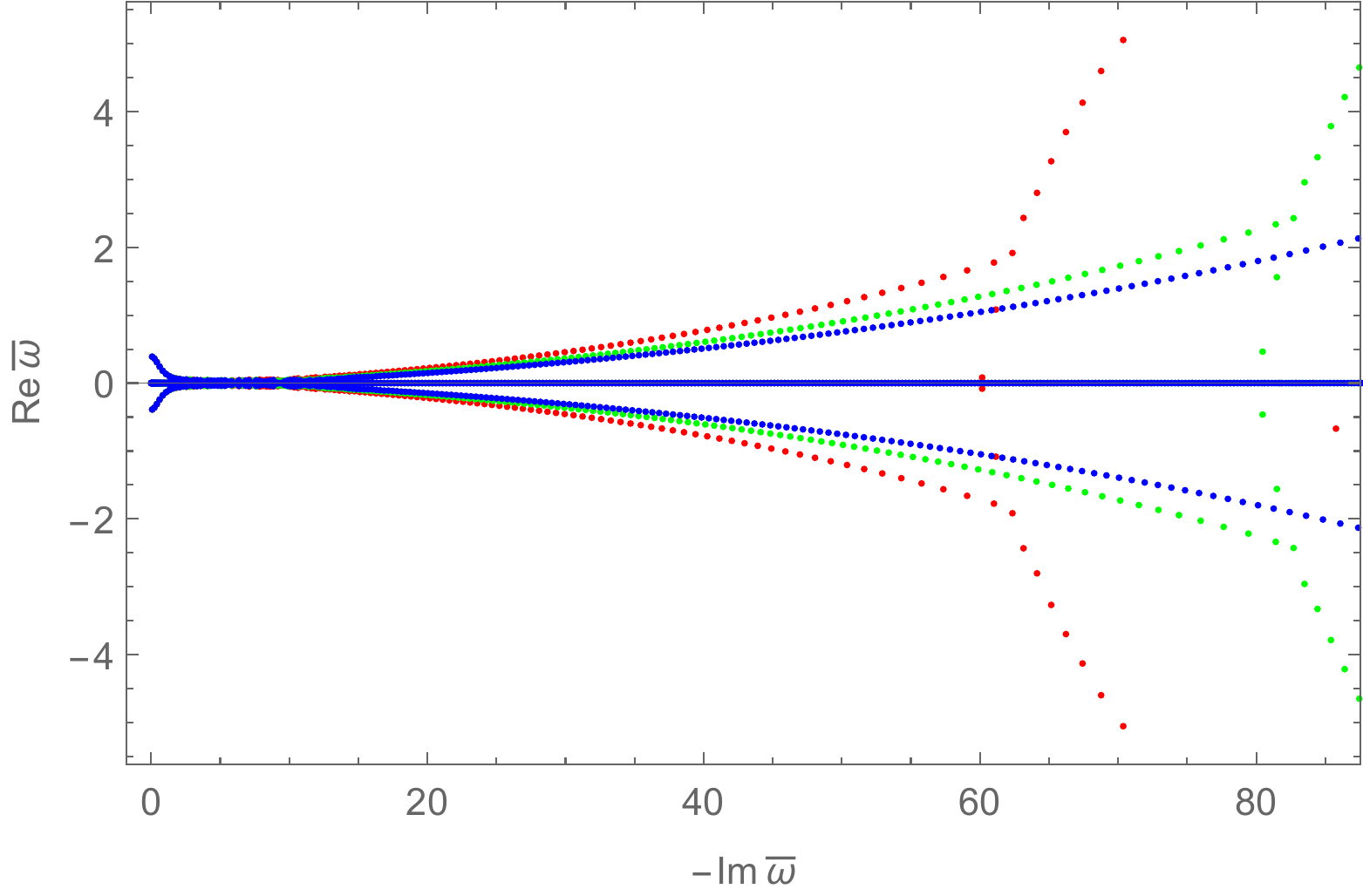}}}
\rotatebox{0}{\resizebox{150pt}{200pt}{\includegraphics[width=0.23\textwidth]{frequ-polar-big-2.pdf}}}\\
\rotatebox{0}{\resizebox{150pt}{200pt}{\includegraphics[width=0.23\textwidth]{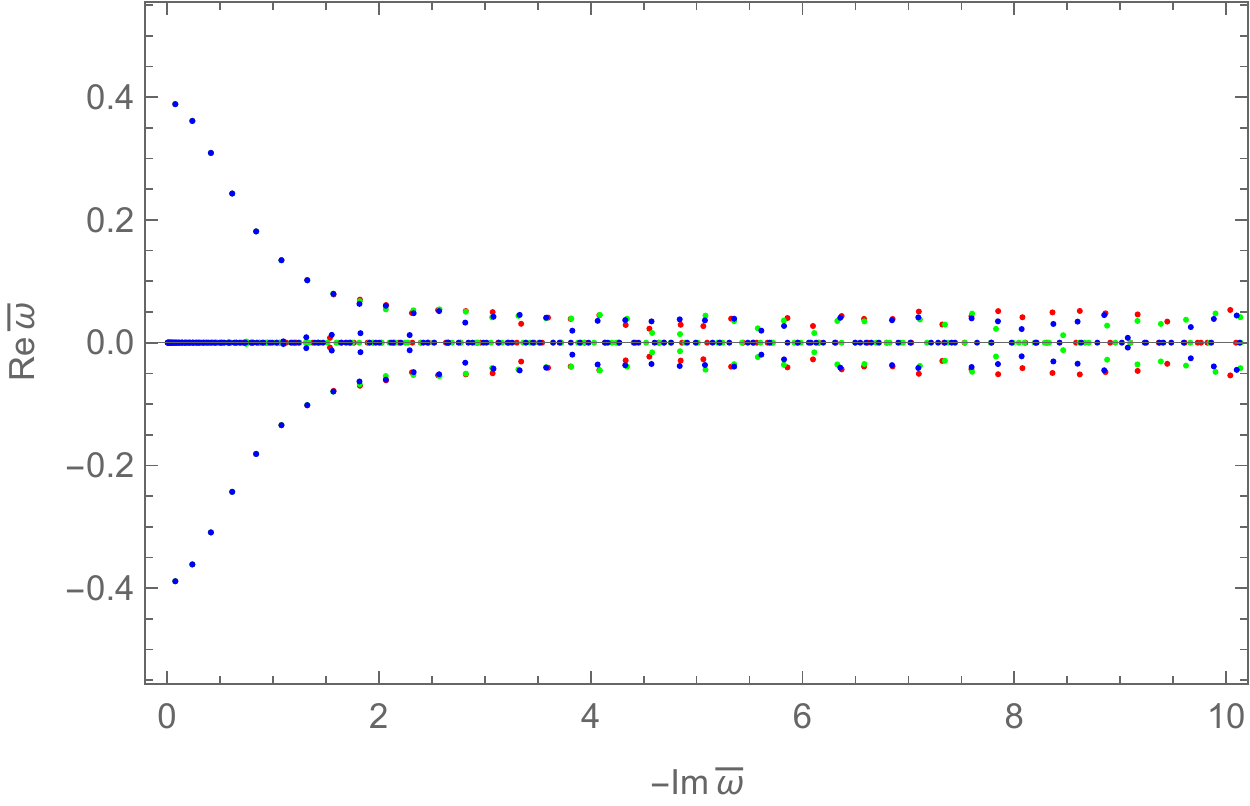}}}
\rotatebox{0}{\resizebox{150pt}{200pt}{\includegraphics[width=0.23\textwidth]{frequ-polar-small-2.pdf}}}\\
\caption{
Comparison of axial to polar modes in pc-GR
(left and right panel, respectively), 
depicting 300 (red dots), 400 (green dots) and
500 (blue dots) iterations. The upper row shows a large range 
in $-\widetilde{\omega}_I$ and the lower row is a 
zoom to the lowest region. Note the structural 
similarity of the axial
to polar modes, suggesting an equivalence,
though not perfect, as in GR.
\label{axial-and-polar}
}
\end{center}
\end{figure}

In Fig. \ref{axial-and-polar} a comparison of axial to
polar modes within pc-GR is shown. 
In the upper row, with a wide range of
$-\widetilde{\omega}_I$, the structure seems to be similar
up to large values of $-\widetilde{\omega}_I$. In the
lower row a Zoom to small values of $-\widetilde{\omega}_I$
is depicted. The $\widetilde{\omega}_R$ polar modes are in 
general larger in pc-GR than in GR.

That there is a branch in the frequency spectrum which
has larger frequencies $\widetilde{\omega}_R$ in pc-GR than
in GR is of importance: The real part of the
frequency is given by
$\widetilde{\omega}_R$ = $m_0 \omega_R$. Using the 
frequency $\nu = 250$Hz and transforming it to units
in km$^{-1}$, one obtains an $\omega_R = 5.24~10^{-3}$km$^{-1}$.
In the first gravitational event observed a mass of the united 
system of about $m_0=60$ solar masses was reported. This gives
a value of $\widetilde{\omega}_R = 0.47$. This is the kind of
order we also obtain. However, he mass $m_0$ was obtained 
recurring to the
GR, i.e., it is {\it theory based}. When in a distinct theory
a larger real frequency $\widetilde{\omega}_R$ is obtained,
keeping $\omega_R$ the same, a larger mass is deduced. This
also implies a larger release in energy and a larger
deduced luminous distance, as suggested in \cite{hess-2016}.

Where the observed distribution of frequencies lies, is a 
matter of
the dynamics of a black hole merger. 
A usual assumption is that all modes can in principle
be excited, but only the low damping modes survive.
In this scenario, no difference between axial and polar modes
are expected and the results will be very similar to GR,
However, when the dynamics permits to excite principally
large damping modes, there might be some hope
to distinguish the theories and clarification can
come from explicit numerical studies.
In case, large damping modes are excited only,
one way to detect a difference is to search, 
for case of large damping modes, for simultaneous 
light events in the same region of the sky where the merger is 
observed. If consistently this light event is at larger
distances than the deduced event, using GR, then it will be
in favor of the existence of additional terms in the metric. 
This depends also on the requirement that a 
light event is produced, requiring some 
mass distribution near to the event,
as an accretions disc.

\section{Conclusions}
\label{concl}

Axial and polar modes where calculated over a wide range
of damping,
within the
{\it General Relativity} (GR) and possible extensions, 
involving a parametric mass-function $m(r)$,
whose leading term correction is proportional to $1/r^4$.
In particular 
the {\it pseudo-complex General Relativity} (pc-GR),
leads to such a particular 
extension of the parameter mass-function.
This mass-function includes a coupling constant of 
the central mass to the dark energy and it was chosen
such that still an event-horizon exists, resulting
in an easier treatment of the QNM, analog to the one
in standard GR. 
The Regge-Wheeler equation for the axial modes and
the Zerilli equation for the polar modes were derived, with
their corresponding potentials. 
After having constructed $\gamma (r)$,
it was shown that
axial and polar modes, though different, still share some
common features. Isospectrality is not maintained, a
feature we still would like to understand and we refer
to future work in progress.
The modes were
found to be stable, implying that the corrections to the
metric lead to consistent results. 

Adding a further $r$-dependence to the mass-function leads
to a branch of frequencies at high damping, resulting
in larger deduced masses than in GR, while for low
damping no large differences are observed.

Assuming that all frequencies are excited, only the
low damping modes survive, resulting in no
detectable differences between pcGR and GR. However,
when the frequencies distribution 
in a merger lies in the large damping region, the deduced masses in the extended version
are larger, implying also
a larger distance to a gravitational wave event. 
If one detects at the same time of this event a light
emission, the two observations result in a different
distance using GR.

The present results serve as a starting point to understand the changes involved in the frequency distribution of
the ring down modes in extending the theory of GR,
which results in a parametric mass-function $m(r)$
in the metric components.

In the Appendices explicit derivation of the ansatz for
$Z^{(+)}$ is given, 
which includes the one proposed 
by S. Chandrasekhar in \cite{chandra,chandra1975a}. 

In a future publication, we will address the pc-Kerr metric,
with however more involved equations. This will pose a problem
to the numerical method used.

\section*{Acknowledgments}
Financial support from DGAPA-PAPIIT (IN100421 and IN114821) is acknowledged.

\section*{Appendix A: Axial potential}
\label{APP-A}

Using $Q=rZ^{(-)}$ gives for the first term in 
(\ref{final-})

\beqa
&
\Delta \frac{d}{dr} \frac{\Delta}{r^4} \frac{dQ}{dr} ~ = ~ \Delta \frac{d}{dr} \frac{\Delta}{r^4}
\left[ r\frac{d}{dr} Z^{(-)} + Z^{(-)}\right]
\nonumber \\
&
~=~
\Delta \frac{d}{dr} \left[ \frac{\Delta}{r^3} \frac{dZ^{(-)}}{dr} + \frac{\Delta}{r^4} Z^{(-)} \right]
&
\nonumber \\
&
~=~
\Delta \left[\frac{d}{dr} \left( \frac{\Delta}{r^4}\right) 
Z^{(-)} + \frac{\Delta}{r^4} \frac{dZ^{(-)}}{dr} \right]
+ \Delta \left[ \frac{1}{r} \frac{\Delta}{r^2}
\frac{dZ^{(-1)}}{dr} \right]
&
\nonumber \\
&
~=~
\Delta \left[\frac{d}{dr} \left( \frac{\Delta}{r^4}\right) 
Z^{(-)} + \frac{\Delta}{r^4} \frac{dZ^{(-)}}{dr} \right]
- \left(\frac{\Delta}{r^2}\right)^2 \frac{dZ^{(-)}}{dr}
+ \frac{\Delta}{r}\frac{d}{dr} \frac{\Delta}{r^2}
\frac{dZ^{(-)}}{dr}
~~~.
\label{eqA1}
\eeqa
Using that $\frac{\Delta}{r^2}\frac{d}{dr} = \frac{d}{dr_*}$, we arrive finally at

\beqa
r\frac{d^2 Z^{(-)}}{dr_*^2} + \Delta \frac{d}{dr} \left( \frac{\Delta}{r^4}\right) Z^{(-)}
~~~.
\label{eqA2}
\eeqa

This has to be substituted into the differential equation (\ref{final-}), leading to

\beqa
r\frac{d^2}{dr_*^2} Z^{(-)} + \Delta \frac{d}{dr} \left( \frac{\Delta}{r^4} \right) Z^{(-)} - \mu^2 \frac{\Delta}{r^3}
Z^{(-)} + r\omega^2 Z^{(-)} & = & 0
~~~.
\label{eqdiff}
\eeqa

Dividing by $r$ and reordering the terms in this 
differential equation, leads to (\ref{zerilli}), with the potential given in 
(\ref{pot-2}).

\section*{Appendix B: Ansatz for the polar mode wave function}
\label{APP-B}

The MATHEMATICA code, used to derive the equations in 
this section, can be retrieved from \cite{matdetails}.

The general ansatz (\ref{eq2.7}) is used, also valid
for a constant mass-function, for which
the expression 
in \cite{chandra,chandra1975b} is recovered.

Using  (\ref{eq2.7}), namely

\beqa
Z^{(+)} (r) & = & \alpha (r) N(r)
+ \beta (r) V(r) + \gamma (r) LX(r)
\label{appB-1}
\eeqa
and applying the operator $\frac{d^2}{dr_*^2}$ to
$Z^{(+)}(r)$, we found \cite{matdetails}
that the factor of $N(r)$ does
not depend on the frequency squared $\omega^2$, only
the factors of $V(r)$ and $LX(r)$ do. We use the 
definitions

\beqa
LX (r) & = & L(r) + X(r)
\nonumber \\
X(r) & = & n V(r)
~~~.
\label{appB-2}
\eeqa 

Concentrating only on the component proportional to
$\omega^2$,
one should obtain for the factor of $V(r)$ and $LX(r)$
the result 
$-\omega^2$ $\left( \beta (r) V(r) + \gamma (r)
LX(r)\right)$. The factor 
obtained after the application of
$\frac{d^2}{dr_*^2}$ onto
$V(r)$ is 
$\left(n \alpha (r) -  \beta (r)\right)$,
which must be equated to $-\beta (r)$. This demands

\beqa
\alpha (r) = 0
~~~.
\label{appB-3}
\eeqa
This automatically reduces (\ref{appB-1}) to
$Z^{(+)} (r) =  \beta (r) V(r) + \gamma (r) LX(r)$.

For the factor of $LX (r)$, restricting to the one
proportional to
$\omega^2$, and using (\ref{appB-3}) we obtain
$-2\beta (r) + n\gamma (r) + 2r\beta^\prime (r)$ = 
$n\gamma (r)$ (the prime refers to the derivative in $r$), 
which leads to the differential equation

\beqa
\beta^\prime & = & \frac{1}{r}\beta
~~~,
\label{appB-4}
\eeqa
with the solution

\beqa
\beta (r) & = & r
~~~.
\label{appB-5}
\eeqa

The $\gamma (r)$ function was determined earlier and 
is given in (\ref{gam-r})
As seen in Appendix C, this solution satisfies the 
equation $G(r)=0$ (\ref{dif-gam}) 
in a wide range of $r$, save near
the point $r=\frac{3}{2}m_0$, implying that the 
Zerilli equation can be approximately constructed, with
a quite small error.

\section*{Appendix C: 
Comparison of $V^{(+)}_V$ with $V^{(+)}_{LX}$,
using (\ref{gam-r}) for $\gamma (r)$
}
\label{APP-C}

\begin{figure}
\begin{center}
\rotatebox{0}{\resizebox{150pt}{150pt}{\includegraphics[width=0.23\textwidth]{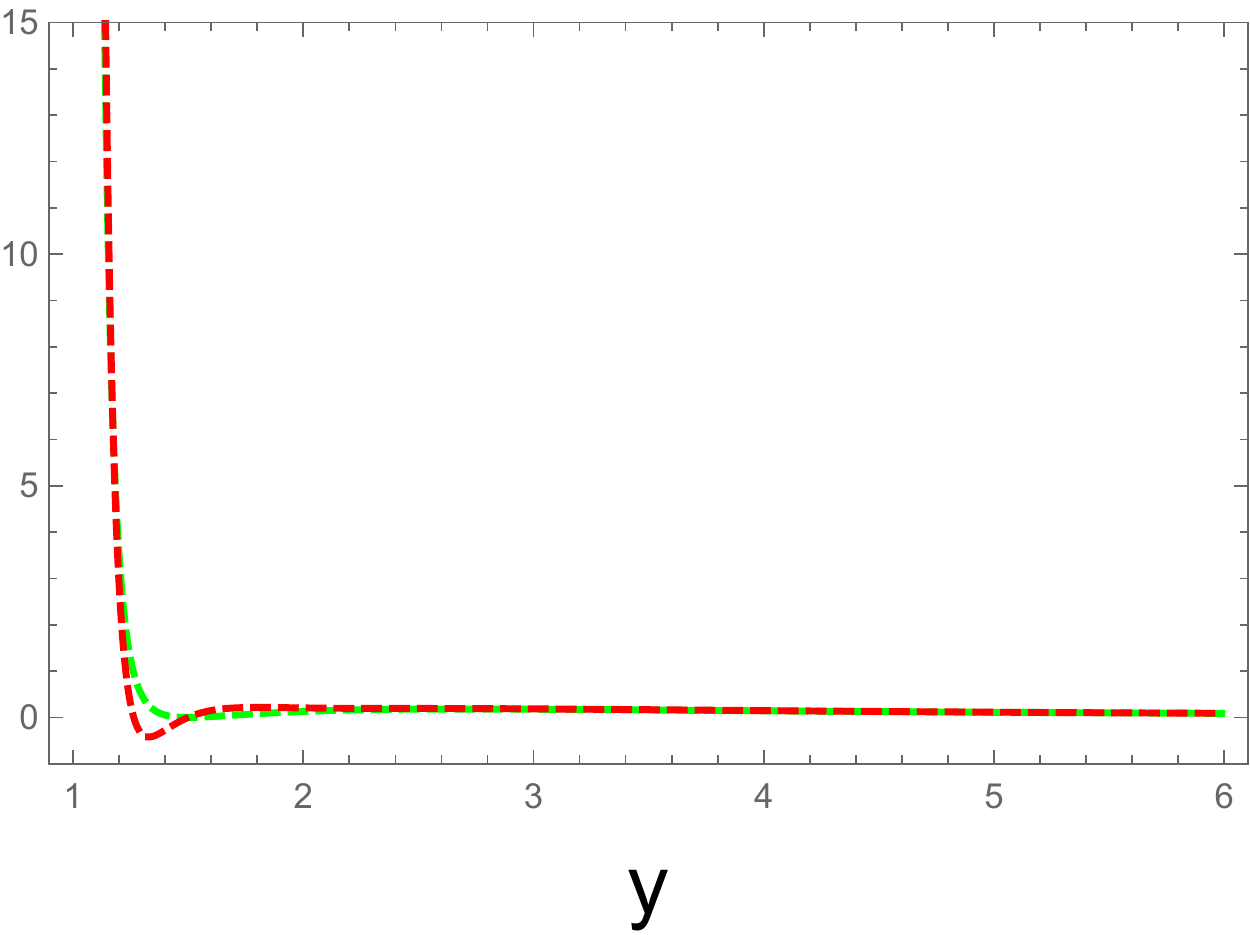}}}
\rotatebox{0}{\resizebox{150pt}{150pt}{\includegraphics[width=0.23\textwidth]{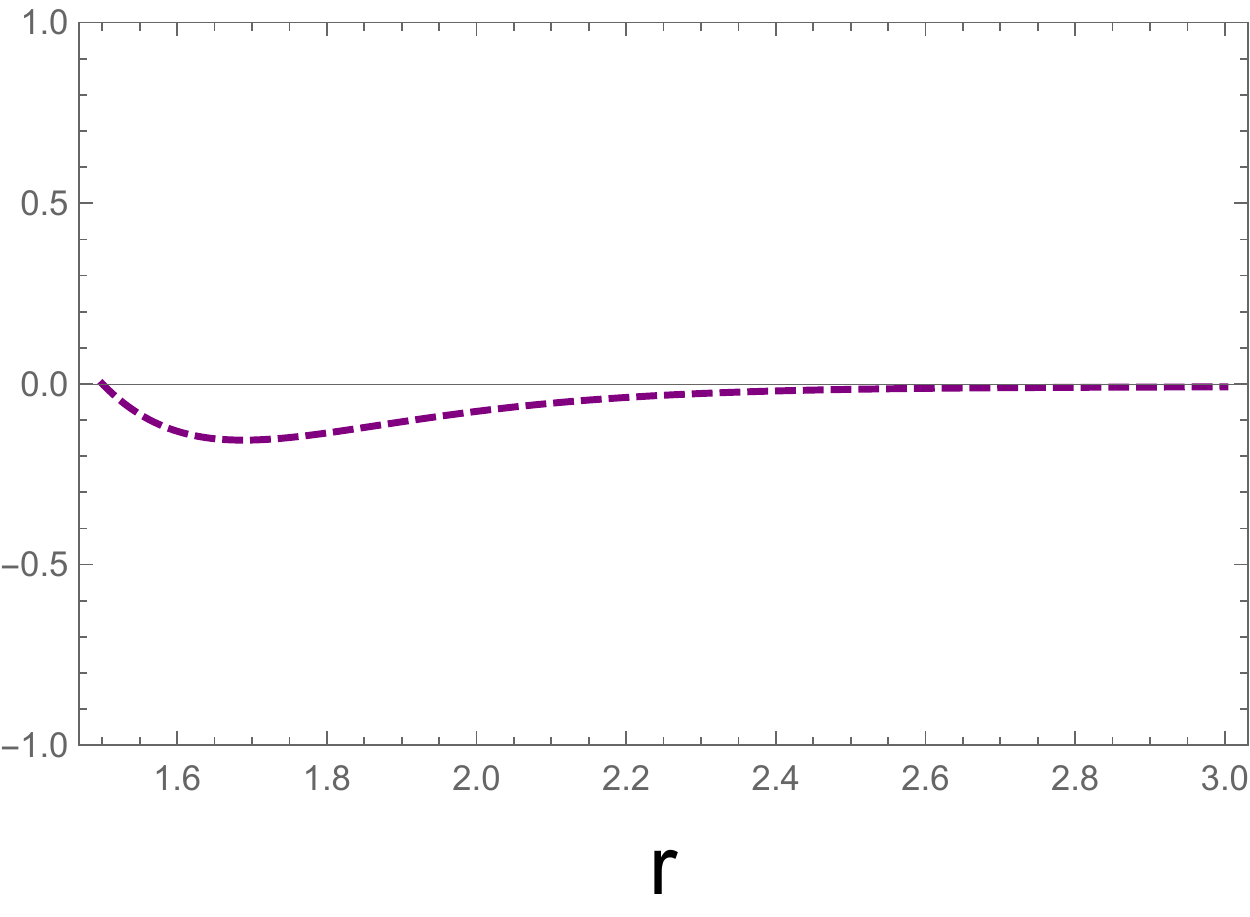}}}
\caption{
The left hand side depicts the two functions $V(r)$ (green
online) and
$VLX(r)$ (red online) in a wide range of $r$. The right
panel shows
the difference $\left(V(r)-VLX(r)\right)$ in a limited 
region in r.
\label{G-error}
}
\end{center}
\end{figure}

In this appendix we analyze the functions
$V^{(+)}(r)$, $V^{(+)}(r)$ and their difference, which
is proportional to the condition
(\ref{dif-gam}).

On the left panel of Fig. \ref{G-error} the two potentials
$V_V^{(+)}$ and $V_{LX}^{(+)}$ are compared for a wide
range of $r$. The potentials were extended to $r<\frac{3}{2}$
in order to appreciate the agreement of both potentials, using
the $\gamma (r)$ as given in (\ref{gam-r}). The two
functions agree very well in a wide range of $r$. 
The only difference appears
near $r=\frac{3}{2}$, which is shown in a reduced scale
on the right hand side of the figure, where, the 
difference $G(r)$ = $\left(V_V^{(+)}-V_{LX}^{(+)}\right)$ 
is depicted.
All in all, the expression for $\gamma (r)$ works very
well, however with the grain of salt that the Zerilli 
equation is not identically satisfied.

\end{document}